%%%%%%%%%%%%%%%%%%%%%%%%%%%%%%%%%%%%%%%%%%%%%%%%

%harvmac
%\input harvmac

%%%%%%%%%%%%%%%%%%  tex macros for preprints, cm version %%%%%%%%%%%%%%
%                     (P. Ginsparg, last updated 9/91)
%                if confused, type `b' in response to query
%
%---------------------------------------------------------------------%
%% site dependent options:
%% \unredoffs and \redoffs define horizontal and vertical offsets
%% respectively for unreduced and reduced modes. \speclscape defines
%% the \special{} call that sets printer to landscape (sideways) mode.
%% from standard set below, leave uncommented as appropriate or redefine
%
%%% next 400dpi
%\def\unredoffs{} \def\redoffs{\voffset=-.31truein\hoffset=-.48truein}
%\def\speclscape{\special{landscape}}
%
%%% apple lw
\def\unredoffs{} \def\redoffs{\voffset=-.31truein\hoffset=-.59truein}
\def\speclscape{\special{ps: landscape}}
%
%%% qms lasergrafix:
%\def\unredoffs{} \def\redoffs{\voffset=-.4truein\hoffset=.125truein}
%\def\speclscape{\special{qms: landscape}}
%
%%% saclay A4 paper:
%\def\unredoffs{\hoffset-.14truein\voffset-.2truein}
%\def\redoffs{\voffset=-.45truein\hoffset=-.21truein}
%\def\speclscape{\special{landscape}}
%
%---------------------------------------------------------------------%
%
\newbox\leftpage \newdimen\fullhsize \newdimen\hstitle \newdimen\hsbody
\tolerance=1000\hfuzz=2pt
\catcode`\@=11 % This allows us to modify PLAIN macros.
\def\bigans{b }
\def\answ{b }

%\message{ big or little (b/l)? }\read-1 to\answ
%
\ifx\answ\bigans\message{(This will come out unreduced.}
\magnification=1200\unredoffs\baselineskip=16pt plus 2pt minus 1pt
\hsbody=\hsize \hstitle=\hsize %take default values for unreduced format
\else\message{(This will be reduced.} \let\l@r=L
\magnification=1000\baselineskip=16pt plus 2pt minus 1pt \vsize=7truein
\redoffs \hstitle=8truein\hsbody=4.75truein\fullhsize=10truein\hsize=\hsbody
\output={\ifnum\pageno=0 %%% This is the HUTP version
  \shipout\vbox{\speclscape{\hsize\fullhsize\makeheadline}
    \hbox to \fullhsize{\hfill\pagebody\hfill}}\advancepageno
  \else
  \almostshipout{\leftline{\vbox{\pagebody\makefootline}}}\advancepageno
  \fi}
\def\almostshipout#1{\if L\l@r \count1=1 \message{[\the\count0.\the\count1]}
      \global\setbox\leftpage=#1 \global\let\l@r=R
 \else \count1=2
  \shipout\vbox{\speclscape{\hsize\fullhsize\makeheadline}
      \hbox to\fullhsize{\box\leftpage\hfil#1}}  \global\let\l@r=L\fi}
\fi
%---------------------------------------------------------------------
%
\newcount\yearltd\yearltd=\year\advance\yearltd by -1900

\def\Title#1#2{\nopagenumbers\abstractfont\hsize=\hstitle\rightline{#1}%
\vskip 1in\centerline{\titlefont #2}\abstractfont\vskip .5in\pageno=0}
\def\Date#1{\vfill\leftline{#1}\tenpoint\supereject\global\hsize=\hsbody%
\footline={\hss\tenrm\folio\hss}}% 	restores pagenumbers
%
%       use following instead of \Date on the preliminary draft,
%       puts date/time on each page in big mode, writes labels in margins

\def\draftmode{\message{ DRAFTMODE }\def\draftdate{{\rm preliminary draft:
\number\month/\number\day/\number\yearltd\ \ \hourmin}}%
\headline={\hfil\draftdate}\writelabels\baselineskip=20pt plus 2pt minus 2pt
 {\count255=\time\divide\count255 by 60 \xdef\hourmin{\number\count255}
  \multiply\count255 by-60\advance\count255 by\time
  \xdef\hourmin{\hourmin:\ifnum\count255<10 0\fi\the\count255}}}
%       use \nolabels to get rid of eqn, ref, and fig labels in draft mode
\def\nolabels{\def\wrlabeL##1{}\def\eqlabeL##1{}\def\reflabeL##1{}}
\def\writelabels{\def\wrlabeL##1{\leavevmode\vadjust{\rlap{\smash%
{\line{{\escapechar=` \hfill\rlap{\sevenrm\hskip.03in\string##1}}}}}}}%
\def\eqlabeL##1{{\escapechar-1\rlap{\sevenrm\hskip.05in\string##1}}}%
\def\reflabeL##1{\noexpand\llap{\noexpand\sevenrm\string\string\string##1}}}
\nolabels
%
% tagged sec numbers
\global\newcount\secno \global\secno=0
\global\newcount\meqno \global\meqno=1
\def\newsec#1{\global\advance\secno by1\message{(\the\secno. #1)}
%\ifx\answ\bigans \vfill\eject \else \bigbreak\bigskip \fi  %if desired
\global\subsecno=0\eqnres@t\noindent{\bf\the\secno. #1}
\writetoca{{\secsym} {#1}}\par\nobreak\medskip\nobreak}
\def\eqnres@t{\xdef\secsym{\the\secno.}\global\meqno=1\bigbreak\bigskip}
\def\sequentialequations{\def\eqnres@t{\bigbreak}}\xdef\secsym{}
\global\newcount\subsecno \global\subsecno=0
\def\subsec#1{\global\advance\subsecno by1\message{(\secsym\the\subsecno. #1)}
\ifnum\lastpenalty>9000\else\bigbreak\fi
\noindent{\it\secsym\the\subsecno. #1}\writetoca{\string\quad
{\secsym\the\subsecno.} {#1}}\par\nobreak\medskip\nobreak}
\def\appendix#1#2{\global\meqno=1\global\subsecno=0\xdef\secsym{\hbox{#1.}}
\bigbreak\bigskip\noindent{\bf Appendix #1. #2}\message{(#1. #2)}
\writetoca{Appendix {#1.} {#2}}\par\nobreak\medskip\nobreak}
%
%       \eqn\label{a+b=c}	gives displayed equation, numbered
%				consecutively within sections.
%     \eqnn and \eqna define labels in advance (of eqalign?)
%
\def\eqnn#1{\xdef #1{(\secsym\the\meqno)}\writedef{#1\leftbracket#1}%
\global\advance\meqno by1\wrlabeL#1}
\def\eqna#1{\xdef #1##1{\hbox{$(\secsym\the\meqno##1)$}}
\writedef{#1\numbersign1\leftbracket#1{\numbersign1}}%
\global\advance\meqno by1\wrlabeL{#1$\{\}$}}
\def\eqn#1#2{\xdef #1{(\secsym\the\meqno)}\writedef{#1\leftbracket#1}%
\global\advance\meqno by1$$#2\eqno#1\eqlabeL#1$$}
%
%			 footnotes
\newskip\footskip\footskip14pt plus 1pt minus 1pt %sets footnote baselineskip
\def\footnotefont{\ninepoint}\def\f@t#1{\footnotefont #1\@foot}
\def\f@@t{\baselineskip\footskip\bgroup\footnotefont\aftergroup\@foot\let\next}
\setbox\strutbox=\hbox{\vrule height9.5pt depth4.5pt width0pt}
\global\newcount\ftno \global\ftno=0
\def\foot{\global\advance\ftno by1\footnote{$^{\the\ftno}$}}
%
%say \footend to put footnotes at end
%will cause problems if \ref used inside \foot, instead use \nref before
\newwrite\ftfile
\def\footend{\def\foot{\global\advance\ftno by1\chardef\wfile=\ftfile
$^{\the\ftno}$\ifnum\ftno=1\immediate\openout\ftfile=foots.tmp\fi%
\immediate\write\ftfile{\noexpand\smallskip%
\noexpand\item{f\the\ftno:\ }\pctsign}\findarg}%
\def\footatend{\vfill\eject\immediate\closeout\ftfile{\parindent=20pt
\centerline{\bf Footnotes}\nobreak\bigskip\input foots.tmp }}}
\def\footatend{}
%
%     \ref\label{text}
% generates a number, assigns it to \label, generates an entry.
% To list the refs on a separate page,  \listrefs
%
\global\newcount\refno \global\refno=1
\newwrite\rfile
\def\ref{[\the\refno]\nref}
\def\nref#1{\xdef#1{[\the\refno]}\writedef{#1\leftbracket#1}%
\ifnum\refno=1\immediate\openout\rfile=refs.tmp\fi
\global\advance\refno by1\chardef\wfile=\rfile\immediate
\write\rfile{\noexpand\item{#1\ }\reflabeL{#1\hskip.31in}\pctsign}\findarg}
%	horrible hack to sidestep tex \write limitation
\def\findarg#1#{\begingroup\obeylines\newlinechar=`\^^M\pass@rg}
{\obeylines\gdef\pass@rg#1{\writ@line\relax #1^^M\hbox{}^^M}%
\gdef\writ@line#1^^M{\expandafter\toks0\expandafter{\striprel@x #1}%
\edef\next{\the\toks0}\ifx\next\em@rk\let\next=\endgroup\else\ifx\next\empty%
\else\immediate\write\wfile{\the\toks0}\fi\let\next=\writ@line\fi\next\relax}}
\def\striprel@x#1{} \def\em@rk{\hbox{}}
\def\lref{\begingroup\obeylines\lr@f}
\def\lr@f#1#2{\gdef#1{\ref#1{#2}}\endgroup\unskip}

\def\addref#1{\immediate\write\rfile{\noexpand\item{}#1}} %now unnecessary
\def\startrefs#1{\immediate\openout\rfile=refs.tmp\refno=#1}
\def\xref{\expandafter\xr@f}\def\xr@f[#1]{#1}
\def\refs#1{\count255=1[\r@fs #1{\hbox{}}]}
\def\r@fs#1{\ifx\und@fined#1\message{reflabel \string#1 is undefined.}%
\nref#1{need to supply reference \string#1.}\fi%
\vphantom{\hphantom{#1}}\edef\next{#1}\ifx\next\em@rk\def\next{}%
\else\ifx\next#1\ifodd\count255\relax\xref#1\count255=0\fi%
\else#1\count255=1\fi\let\next=\r@fs\fi\next}
%

%
% this is ugly, but moore insists
\newwrite\ffile\global\newcount\figno \global\figno=1
\def\fig{fig.~\the\figno\nfig}
\def\nfig#1{\xdef#1{fig.~\the\figno}%
\writedef{#1\leftbracket fig.\noexpand~\the\figno}%
\ifnum\figno=1\immediate\openout\ffile=figs.tmp\fi\chardef\wfile=\ffile%
\immediate\write\ffile{\noexpand\medskip\noexpand\item{Fig.\ \the\figno. }
\reflabeL{#1\hskip.55in}\pctsign}\global\advance\figno by1\findarg}
\def\xfig{\expandafter\xf@g}\def\xf@g fig.\penalty\@M\ {}
\def\figs#1{figs.~\f@gs #1{\hbox{}}}
\def\f@gs#1{\edef\next{#1}\ifx\next\em@rk\def\next{}\else
\ifx\next#1\xfig #1\else#1\fi\let\next=\f@gs\fi\next}
\newwrite\lfile
{\escapechar-1\xdef\pctsign{\string\%}\xdef\leftbracket{\string\{}
\xdef\rightbracket{\string\}}\xdef\numbersign{\string\#}}

\def\writestop{\def\writestoppt{\immediate\write\lfile{\string\pageno%
\the\pageno\string\startrefs\leftbracket\the\refno\rightbracket%
\string\def\string\secsym\leftbracket\secsym\rightbracket%
\string\secno\the\secno\string\meqno\the\meqno}\immediate\closeout\lfile}}
\def\writestoppt{}\def\writedef#1{}
\def\seclab#1{\xdef #1{\the\secno}\writedef{#1\leftbracket#1}\wrlabeL{#1=#1}}
\def\subseclab#1{\xdef #1{\secsym\the\subsecno}%
\writedef{#1\leftbracket#1}\wrlabeL{#1=#1}}
\newwrite\tfile \def\writetoca#1{}
\def\leaderfill{\leaders\hbox to 1em{\hss.\hss}\hfill}
%	use this to write file with table of contents
\def\writetoc{\immediate\openout\tfile=toc.tmp
   \def\writetoca##1{{\edef\next{\write\tfile{\noindent ##1
   \string\leaderfill {\noexpand\number\pageno} \par}}\next}}}
%       and this lists table of contents on second pass

%
\catcode`\@=12 % at signs are no longer letters
%
%	Unpleasantness in calling in abstract and title fonts
\edef\tfontsize{\ifx\answ\bigans scaled\magstep3\else scaled\magstep4\fi}
\font\titlerm=cmr10 \tfontsize \font\titlerms=cmr7 \tfontsize
\font\titlermss=cmr5 \tfontsize \font\titlei=cmmi10 \tfontsize
\font\titleis=cmmi7 \tfontsize \font\titleiss=cmmi5 \tfontsize
\font\titlesy=cmsy10 \tfontsize \font\titlesys=cmsy7 \tfontsize
\font\titlesyss=cmsy5 \tfontsize \font\titleit=cmti10 \tfontsize
\skewchar\titlei='177 \skewchar\titleis='177 \skewchar\titleiss='177
\skewchar\titlesy='60 \skewchar\titlesys='60 \skewchar\titlesyss='60
\def\titlefont{\def\rm{\fam0\titlerm}% switch to title font
\textfont0=\titlerm \scriptfont0=\titlerms \scriptscriptfont0=\titlermss
\textfont1=\titlei \scriptfont1=\titleis \scriptscriptfont1=\titleiss
\textfont2=\titlesy \scriptfont2=\titlesys \scriptscriptfont2=\titlesyss
\textfont\itfam=\titleit \def\it{\fam\itfam\titleit}\rm}
 \ifx\answ\bigans\else scaled\magstep1\fi
\ifx\answ\bigans\def\abstractfont{\tenpoint}\else
\font\abssl=cmsl10 scaled \magstep1
\font\absrm=cmr10 scaled\magstep1 \font\absrms=cmr7 scaled\magstep1
\font\absrmss=cmr5 scaled\magstep1 \font\absi=cmmi10 scaled\magstep1
\font\absis=cmmi7 scaled\magstep1 \font\absiss=cmmi5 scaled\magstep1
\font\abssy=cmsy10 scaled\magstep1 \font\abssys=cmsy7 scaled\magstep1
\font\abssyss=cmsy5 scaled\magstep1 \font\absbf=cmbx10 scaled\magstep1
\skewchar\absi='177 \skewchar\absis='177 \skewchar\absiss='177
\skewchar\abssy='60 \skewchar\abssys='60 \skewchar\abssyss='60
\def\abstractfont{\def\rm{\fam0\absrm}% switch to abstract font
\textfont0=\absrm \scriptfont0=\absrms \scriptscriptfont0=\absrmss
\textfont1=\absi \scriptfont1=\absis \scriptscriptfont1=\absiss
\textfont2=\abssy \scriptfont2=\abssys \scriptscriptfont2=\abssyss
\textfont\itfam=\bigit \def\it{\fam\itfam\bigit}\def\footnotefont{\tenpoint}%
\textfont\slfam=\abssl \def\sl{\fam\slfam\abssl}%
\textfont\bffam=\absbf \def\bf{\fam\bffam\absbf}\rm}\fi
\def\tenpoint{\def\rm{\fam0\tenrm}% switch back to 10-point type
\textfont0=\tenrm \scriptfont0=\sevenrm \scriptscriptfont0=\fiverm
\textfont1=\teni  \scriptfont1=\seveni  \scriptscriptfont1=\fivei
\textfont2=\tensy \scriptfont2=\sevensy \scriptscriptfont2=\fivesy
\textfont\itfam=\tenit \def\it{\fam\itfam\tenit}\def\footnotefont{\ninepoint}%
\textfont\bffam=\tenbf \def\bf{\fam\bffam\tenbf}\def\sl{\fam\slfam\tensl}\rm}
\font\ninerm=cmr9 \font\sixrm=cmr6 \font\ninei=cmmi9 \font\sixi=cmmi6
\font\ninesy=cmsy9 \font\sixsy=cmsy6 \font\ninebf=cmbx9
\font\nineit=cmti9 \font\ninesl=cmsl9 \skewchar\ninei='177
\skewchar\sixi='177 \skewchar\ninesy='60 \skewchar\sixsy='60
\def\ninepoint{\def\rm{\fam0\ninerm}% switch to footnote font
\textfont0=\ninerm \scriptfont0=\sixrm \scriptscriptfont0=\fiverm
\textfont1=\ninei \scriptfont1=\sixi \scriptscriptfont1=\fivei
\textfont2=\ninesy \scriptfont2=\sixsy \scriptscriptfont2=\fivesy
\textfont\itfam=\ninei \def\it{\fam\itfam\nineit}\def\sl{\fam\slfam\ninesl}%
\textfont\bffam=\ninebf \def\bf{\fam\bffam\ninebf}\rm}
%
%---------------------------------------------------------------------
%

\hyphenation{anom-aly anom-alies coun-ter-term coun-ter-terms}
\def\inv{^{\raise.15ex\hbox{${\scriptscriptstyle -}$}\kern-.05em 1}}

\def\Dsl{\,\raise.15ex\hbox{/}\mkern-13.5mu D} %this one can be subscripted
\def\dsl{\raise.15ex\hbox{/}\kern-.57em\partial}

\font\bigit=cmti10 scaled \magstep1
 %pound sterling
\def\lspace{\ifx\answ\bigans{}\else\qquad\fi}
\def\lbspace{\ifx\answ\bigans{}\else\hskip-.2in\fi} % $$\lbspace...$$
\def\boxeqn#1{\vcenter{\vbox{\hrule\hbox{\vrule\kern3pt\vbox{\kern3pt
	\hbox{${\displaystyle #1}$}\kern3pt}\kern3pt\vrule}\hrule}}}
\def\mbox#1#2{\vcenter{\hrule \hbox{\vrule height#2in
		\kern#1in \vrule} \hrule}}  %e.g. \mbox{.1}{.1}
%	matters of taste
%\def\tilde{\widetilde} \def\bar{\overline} \def\hat{\widehat}
%
% some sample definitions
  %     curly letters
 \def\CC{{\cal C}}

\def\vev#1{\langle #1 \rangle}

\def\darr#1{\raise1.5ex\hbox{$\leftrightarrow$}\mkern-16.5mu #1}
 %pound sterling

 %puts a small half in a displayed eqn
\def\roughly#1{\raise.3ex\hbox{$#1$\kern-.75em\lower1ex\hbox{$\sim$}}}

%\draftmode
\let\includefigures=\iftrue
\let\useblackboard=\iftrue
\newfam\black

%Figure Stuff
\includefigures
\message{If you do not have epsf.tex (to include figures),}
\message{change the option at the top of the tex file.}
\input epsf
\def\figin{\epsfcheck\figin}\def\figins{\epsfcheck\figins}
\def\epsfcheck{\ifx\epsfbox\UnDeFiNeD
\message{(NO epsf.tex, FIGURES WILL BE IGNORED)}
\gdef\figin##1{\vskip2in}\gdef\figins##1{\hskip.5in}% blank space instead
\else\message{(FIGURES WILL BE INCLUDED)}%
\gdef\figin##1{##1}\gdef\figins##1{##1}\fi}
\def\DefWarn#1{}
\def\figinsert{\goodbreak\midinsert}
\def\ifig#1#2#3{\DefWarn#1\xdef#1{fig.~\the\figno}
\writedef{#1\leftbracket fig.\noexpand~\the\figno}%
\figinsert\figin{\centerline{#3}}\medskip\centerline{\vbox{
\baselineskip12pt\advance\hsize by -1truein
\noindent\footnotefont{\bf Fig.~\the\figno:} #2}}
%\bigskip
\endinsert\global\advance\figno by1}
%%%
\else
\def\ifig#1#2#3{\xdef#1{fig.~\the\figno}
\writedef{#1\leftbracket fig.\noexpand~\the\figno}%
%\figinsert\figin{\centerline{#3}}\medskip
%\centerline{\vbox{\baselineskip12pt
%\advance\hsize by -1truein\noindent
%\footnotefont{\bf Fig.~\the\figno:} #2}}
%\bigskip\endinsert
\global\advance\figno by1} \fi

\def\id{{1 \kern-.28em {\rm l}}}

\def\K3{{\bf K3}}
\def\journal#1&#2(#3){\unskip, \sl #1\ \bf #2 \rm(19#3) }
\def\andjournal#1&#2(#3){\sl #1~\bf #2 \rm (19#3) }

\def\hat{\widehat}
\def\ie{{\it i.e.}}

\def\tilde{\widetilde}

\def\frac#1#2{{#1\over#2}}

\def\vev#1{\langle#1\rangle}

\def\inbar{\,\vrule height1.5ex width.4pt depth0pt}
\def\IC{\relax\hbox{$\inbar\kern-.3em{\rm C}$}}
\def\IR{\relax{\rm I\kern-.18em R}}
\def\IP{\relax{\rm I\kern-.18em P}}

%
%%%%%%%%%%%%%%%%%%%%%%%%%%%%%%%%%%%%
%

%
\catcode`\@=11
\def\slash#1{\mathord{\mathpalette\c@ncel{#1}}}
\overfullrule=0pt
\def\AA{{\cal A}}
\def\BB{{\cal B}}
\def\CC{{\cal C}}

\def\LL{{\cal L}}
\def\NN{{\cal N}}
\def\OO{{\cal O}}

\def\underrel#1\over#2{\mathrel{\mathop{\kern\z@#1}\limits_{#2}}}

\catcode`\@=12

%%%%%%%%%%%%%%%%%%%%%%%%%%%%%%%%%%%%%%%%%%%%%%%%%%%%%%%%%%%%%%

%

\def\vev#1{\left\langle #1 \right\rangle}
\def\det{{\rm det}}

\def\det{{\rm det}}
\def\exp{{\rm exp}}

%%%%%%%%%%%%%%%%%%%%%%%%%%%%%%%%%%%%%%%%%%%%%%%%%%%%%%%%%%%%%%
% new defs:

\def\lambdah{{\hat{\lambda}}}

\def\p{{\partial}}

\def\ra{{\rightarrow}}

\def\tq{{\tilde{q}}}
\def\tom{{\tilde{\omega}}}

\def\LL{{\cal L}}
\def\tf{{\tilde f}}
\def\tR{{\tilde R}}
\def\tQ{{\tilde Q}}
\def\tgamma{{\tilde \gamma}}
\def\tT{{\tilde T}}

%Central Charges and higher derivative terms

%\cite{Metsaev:1986yb}
\lref\MetsaevTseytlina{
  R.~R.~Metsaev and A.~A.~Tseytlin,
  ``Curvature Cubed Terms in String Theory Effective Actions,''
  Phys.\ Lett.\  B {\bf 185}, 52 (1987).
  %%CITATION = PHLTA,B185,52;%%
}

%\cite{Bastianelli:1999ab}
\lref\BastianelliFTa{
  F.~Bastianelli, S.~Frolov and A.~A.~Tseytlin,
  ``Three-point correlators of stress tensors in maximally-supersymmetric
  conformal theories in d = 3 and d = 6,''
  Nucl.\ Phys.\  B {\bf 578}, 139 (2000)
  [arXiv:hep-th/9911135].
  %%CITATION = NUPHA,B578,139;%%
}

%\cite{Bastianelli:2000hi}
\lref\BastianelliFTb{
  F.~Bastianelli, S.~Frolov and A.~A.~Tseytlin,
  ``Conformal anomaly of (2,0) tensor multiplet in six dimensions and  AdS/CFT
  correspondence,''
  JHEP {\bf 0002}, 013 (2000)
  [arXiv:hep-th/0001041].
  %%CITATION = JHEPA,0002,013;%%
}

%Causality violation in higher derivative gravity

%\cite{Buchel:2009tt}
\lref\BuchelMyers{
  A.~Buchel and R.~C.~Myers,
  ``Causality of Holographic Hydrodynamics,''
  arXiv:0906.2922 [hep-th].
  %%CITATION = ARXIV:0906.2922;%%
}

%\cite{Brigante:2007nu}
\lref\BriganteLMSYa{
  M.~Brigante, H.~Liu, R.~C.~Myers, S.~Shenker and S.~Yaida,
  ``Viscosity Bound Violation in Higher Derivative Gravity,''
  Phys.\ Rev.\  D {\bf 77}, 126006 (2008)
  [arXiv:0712.0805 [hep-th]].
  %%CITATION = PHRVA,D77,126006;%%
}

%\cite{Brigante:2008gz}
\lref\BriganteLMSYb{
  M.~Brigante, H.~Liu, R.~C.~Myers, S.~Shenker and S.~Yaida,
  ``The Viscosity Bound and Causality Violation,''
  Phys.\ Rev.\ Lett.\  {\bf 100}, 191601 (2008)
  [arXiv:0802.3318 [hep-th]].
  %%CITATION = PRLTA,100,191601;%%
}

% Bound on the Central Charges

%\cite{Hofman:2009ug}
\lref\Hofman{
  D.~M.~Hofman,
  ``Higher Derivative Gravity, Causality and Positivity of Energy in a UV
  complete QFT,''
  arXiv:0907.1625 [hep-th].
  %%CITATION = ARXIV:0907.1625;%%
}

%\cite{Hofman:2008ar}
\lref\HofmanMaldacena{
  D.~M.~Hofman and J.~Maldacena,
  ``Conformal collider physics: Energy and charge correlations,''
  JHEP {\bf 0805}, 012 (2008)
  [arXiv:0803.1467 [hep-th]].
  %%CITATION = JHEPA,0805,012;%%
}

% Lovelock Gravity Black Holes

%\cite{Dehghani:2009zz}
\lref\DehghaniPourhasan{
  M.~H.~Dehghani and R.~Pourhasan,
  ``Thermodynamic instability of black holes of third order Lovelock gravity,''
  Phys.\ Rev.\  D {\bf 79}, 064015 (2009)
  [arXiv:0903.4260 [gr-qc]].
  %%CITATION = PHRVA,D79,064015;%%
}

%\cite{Dehghani:2008ye}
\lref\DehghaniBH{
  M.~H.~Dehghani, N.~Bostani and S.~H.~Hendi,
  ``Magnetic Branes in Third Order Lovelock-Born-Infeld Gravity,''
  Phys.\ Rev.\  D {\bf 78}, 064031 (2008)
  [arXiv:0806.1429 [gr-qc]].
  %%CITATION = PHRVA,D78,064031;%%
}

%\cite{Dehghani:2008qr}
\lref\DehghaniAH{
  M.~H.~Dehghani, N.~Alinejadi and S.~H.~Hendi,
  ``Topological Black Holes in Lovelock-Born-Infeld Gravity,''
  Phys.\ Rev.\  D {\bf 77}, 104025 (2008)
  [arXiv:0802.2637 [hep-th]].
  %%CITATION = PHRVA,D77,104025;%%
}

\lref\Lovelock{
  D.~Lovelock
  ``The Einstein Tensor and Its Generalizations''
  J.Math.Phys. \ {\bf 12}, 498 (1971)
}

\lref\Caia{
  R.~G.~Cai,
  ``Gauss-Bonnet black holes in AdS spaces,''
  Phys.\ Rev.\  D {\bf 65}, 084014 (2002)
  [arXiv:hep-th/0109133].
  %%CITATION = PHRVA,D65,084014;%%
}

\lref\Caib{
  R.~G.~Cai, L.~M.~Cao, Y.~P.~Hu and S.~P.~Kim,
  ``Generalized Vaidya Spacetime in Lovelock Gravity and Thermodynamics on
  Apparent Horizon,''
  Phys.\ Rev.\  D {\bf 78}, 124012 (2008)
  [arXiv:0810.2610 [hep-th]].
  %%CITATION = PHRVA,D78,124012;%%
}

\lref\Wheeler{
  J.~T.~Wheeler,
  ``Symmetric Solutions To The Gauss-Bonnet Extended Einstein Equations,''
  Nucl.\ Phys.\  B {\bf 268}, 737 (1986).
  %%CITATION = NUPHA,B268,737;%%

}

\lref\MyersSimon{
  R.~C.~Myers and J.~Z.~Simon,
  ``Black Hole Thermodynamics in Lovelock Gravity,''
  Phys.\ Rev.\  D {\bf 38}, 2434 (1988).
  %%CITATION = PHRVA,D38,2434;%%
}

%\deBoerPN
\lref\deBoerPN{
  J.~de Boer, M.~Kulaxizi and A.~Parnachev,
  ``$AdS_7/CFT_6$, Gauss-Bonnet Gravity, and Viscosity Bound,''
  arXiv:0910.5347 [hep-th].
  %%CITATION = ARXIV:0910.5347;%%
}

\lref\ExirifardJabbari{
  Q.~Exirifard and M.~M.~Sheikh-Jabbari,
  ``Lovelock Gravity at the Crossroads of Palatini and Metric Formulations,''
  Phys.\ Lett.\  B {\bf 661}, 158 (2008)
  [arXiv:0705.1879 [hep-th]].
  %%CITATION = PHLTA,B661,158;%%
}

%\NojiriMH
\lref\NojiriMH{
  S.~Nojiri and S.~D.~Odintsov,
  ``On the conformal anomaly from higher derivative gravity in AdS/CFT
  correspondence,''
  Int.\ J.\ Mod.\ Phys.\  A {\bf 15}, 413 (2000)
  [arXiv:hep-th/9903033].
  %%CITATION = IMPAE,A15,413;%%
}

\lref\HS{
  M.~Henningson and K.~Skenderis,
  ``The holographic Weyl anomaly,''
  JHEP {\bf 9807}, 023 (1998)
  [arXiv:hep-th/9806087];
  %%CITATION = JHEPA,9807,023;%%
%\HenningsonEY
%\lref\HenningsonEY{
%  M.~Henningson and K.~Skenderis,
  ``Holography and the Weyl anomaly,''
  Fortsch.\ Phys.\  {\bf 48}, 125 (2000)
  [arXiv:hep-th/9812032].
  %%CITATION = FPYKA,48,125;%%
}

%\CamanhoVW
\lref\CamanhoVW{
  X.~O.~Camanho and J.~D.~Edelstein,
  ``Causality constraints in AdS/CFT from conformal collider physics and
  Gauss-Bonnet gravity,''
  arXiv:0911.3160 [hep-th].
  %%CITATION = ARXIV:0911.3160;%%
}

%\BuchelSK
\lref\BuchelSK{
  A.~Buchel, J.~Escobedo, R.~C.~Myers, M.~F.~Paulos, A.~Sinha and M.~Smolkin,
  ``Holographic GB gravity in arbitrary dimensions,''
  arXiv:0911.4257 [hep-th].
  %%CITATION = ARXIV:0911.4257;%%
}

%AdS-CFT refs:

\lref\Maldacena{
  J.~M.~Maldacena,
  ``The large N limit of superconformal field theories and supergravity,''
  Adv.\ Theor.\ Math.\ Phys.\  {\bf 2}, 231 (1998)
  [Int.\ J.\ Theor.\ Phys.\  {\bf 38}, 1113 (1999)]
  [arXiv:hep-th/9711200].
  %%CITATION = IJTPB,38,1113;%%
}

\lref\Witten{
  E.~Witten,
  ``Anti-de Sitter space and holography,''
  Adv.\ Theor.\ Math.\ Phys.\  {\bf 2}, 253 (1998)
  [arXiv:hep-th/9802150].
  %%CITATION = 00203,2,253;%%
}

\lref\GKP{
  S.~S.~Gubser, I.~R.~Klebanov and A.~M.~Polyakov,
  ``Gauge theory correlators from non-critical string theory,''
  Phys.\ Lett.\  B {\bf 428}, 105 (1998)
  [arXiv:hep-th/9802109].
  %%CITATION = PHLTA,B428,105;%%
}

\lref\MetsaevTseytlin{
  R.~R.~Metsaev and A.~A.~Tseytlin,
  ``CURVATURE CUBED TERMS IN STRING THEORY EFFECTIVE ACTIONS,''
  Phys.\ Lett.\  B {\bf 185}, 52 (1987).
  %%CITATION = PHLTA,B185,52;%%
}

%\AharonyXZ
\lref\AharonyXZ{
  O.~Aharony, A.~Fayyazuddin and J.~M.~Maldacena,
  ``The large N limit of N = 2,1 field theories from three-branes in
  F-theory,''
  JHEP {\bf 9807}, 013 (1998)
  [arXiv:hep-th/9806159].
  %%CITATION = JHEPA,9807,013;%%
}

%\FayyazuddinFB
\lref\FayyazuddinFB{
  A.~Fayyazuddin and M.~Spalinski,
  ``Large N superconformal gauge theories and supergravity orientifolds,''
  Nucl.\ Phys.\  B {\bf 535}, 219 (1998)
  [arXiv:hep-th/9805096].
  %%CITATION = NUPHA,B535,219;%%
}

%\KatsMQ
\lref\KatsMQ{
  Y.~Kats and P.~Petrov,
  ``Effect of curvature squared corrections in AdS on the viscosity of the dual
  gauge theory,''
  JHEP {\bf 0901}, 044 (2009)
  [arXiv:0712.0743 [hep-th]].
  %%CITATION = JHEPA,0901,044;%%
}

%\BoulwareWK
\lref\BoulwareWK{
  D.~G.~Boulware and S.~Deser,
  ``String Generated Gravity Models,''
  Phys.\ Rev.\ Lett.\  {\bf 55}, 2656 (1985).
  %%CITATION = PRLTA,55,2656;%%
}

%\ZwiebachUQ
\lref\ZwiebachUQ{
  B.~Zwiebach,
  ``Curvature Squared Terms And String Theories,''
  Phys.\ Lett.\  B {\bf 156}, 315 (1985).
  %%CITATION = PHLTA,B156,315;%%
}

%\ZuminoDP
\lref\ZuminoDP{
  B.~Zumino,
  ``Gravity Theories In More Than Four-Dimensions,''
  Phys.\ Rept.\  {\bf 137}, 109 (1986).
  %%CITATION = PRPLC,137,109;%%
}

\lref\KSS{
  P.~Kovtun, D.~T.~Son and A.~O.~Starinets,
  ``Viscosity in strongly interacting quantum field theories from black hole
 physics,''
  Phys.\ Rev.\ Lett.\  {\bf 94}, 111601 (2005)
  [arXiv:hep-th/0405231].
  %%CITATION = PRLTA,94,111601;%%
  ``Holography and hydrodynamics: Diffusion on stretched horizons,''
  JHEP {\bf 0310}, 064 (2003)
  [arXiv:hep-th/0309213].
  %%CITATION = JHEPA,0310,064;%%
}

%\ShuAX
\lref\ShuAX{
  F.~W.~Shu,
  ``The Quantum Viscosity Bound In Lovelock Gravity,''
  arXiv:0910.0607 [hep-th].
  %%CITATION = ARXIV:0910.0607;%%
}

%\HawkingDH
\lref\HawkingDH{
  S.~W.~Hawking and D.~N.~Page,
  ``Thermodynamics Of Black Holes In Anti-De Sitter Space,''
  Commun.\ Math.\ Phys.\  {\bf 87}, 577 (1983).
  %%CITATION = CMPHA,87,577;%%
}

%\WittenZW
\lref\WittenZW{
  E.~Witten,
  ``Anti-de Sitter space, thermal phase transition, and confinement in  gauge
  theories,''
  Adv.\ Theor.\ Math.\ Phys.\  {\bf 2}, 505 (1998)
  [arXiv:hep-th/9803131].
  %%CITATION = 00203,2,505;%%
}

\lref\CE{
X.~O.~Camanho and J.~D.~Edelstein,
  ``Causality in AdS/CFT and Lovelock Theory,''
  to appear.
}

\lref\SonSS{
    D.~T.~Son and A.~O.~Starinets,
  ``Minkowski-space correlators in AdS/CFT correspondence: Recipe and applications,''
    JHEP {\bf 0209}, 042 (2002)
   [arXiv:hep-th/0205051].
   %%CITATION = JHEPA,0209,042;%%
}

\lref\Marolf{
D.~Marolf,
  ``States and boundary terms: Subtleties of Lorentzian AdS/CFT,''
  JHEP {\bf 0505}, 042 (2005)
  [arXiv:hep-th/0412032].
  %%CITATION = JHEPA,0505,042;%%
}

\lref\SkenderisSVR{
   K.~Skenderis and B.~C.~van Rees,
   ``Real-time gauge/gravity duality,''
   [arXiv:hep-th/0805.0150].
   %%CITATION = ARXIV:0805.0150;%%
}

\lref\HerzogHS{
C.~P.~Herzog and D.~T.~Son,
  ``Schwinger-Keldysh propagators from AdS/CFT correspondence,''
   JHEP {\bf 0303}, 046 (2003)
   [arXiv:hep-th/0212072].
 %%CITATION = JHEPA,0303,046;%%
}

\lref\Takahashia{
 T.~Takahashi and J.~Soda,
 ``Stability of Lovelock Black Holes under Tensor Perturbations,''
 Phys.\ Rev.\  D {\bf 79}, 104025 (2009)
 [arXiv:0902.2921 [gr-qc]].
%%CITATION = PHRVA,D79,104025;%%
}

\lref\Takahashib{
  T.~Takahashi and J.~Soda,
 ``Instability of Small Lovelock Black Holes in Even-dimensions,''
  Phys.\ Rev.\  D {\bf 80}, 104021 (2009)
   [arXiv:0907.0556 [gr-qc]].
  %%CITATION = PHRVA,D80,104021;%%
}

\lref\Kofinasa{
  G.~Kofinas and R.~Olea,
  ``Universal Kounterterms in Lovelock AdS gravity,''
  Fortsch.\ Phys.\  {\bf 56}, 957 (2008)
  [arXiv:0806.1197 [hep-th]].
  %%CITATION = FPYKA,56,957;%%
}

\lref\Kofinasb{
  G.~Kofinas and R.~Olea,
  ``Universal regularization prescription for Lovelock AdS gravity,''
  JHEP {\bf 0711}, 069 (2007)
  [arXiv:0708.0782 [hep-th]].
  %%CITATION = JHEPA,0711,069;%%
}

\lref\Oleab{O.~Miskovic and R.~Olea,
  ``Counterterms in Dimensionally Continued AdS Gravity,''
  JHEP {\bf 0710} (2007) 028
  [arXiv:0706.4460 [hep-th]].
  %%CITATION = JHEPA,0710,028;%%
}

\lref\Oleaa{
  G.~Kofinas and R.~Olea,
  ``Vacuum energy in Einstein-Gauss-Bonnet AdS gravity,''
  Phys.\ Rev.\  D {\bf 74}, 084035 (2006)
  [arXiv:hep-th/0606253].
  %%CITATION = PHRVA,D74,084035;%%
}

\lref\Medved{
  R.~Brustein and A.~J.~M.~Medved,
  ``The ratio of shear viscosity to entropy density in generalized theories of
  gravity,''
  Phys.\ Rev.\  D {\bf 79}, 021901 (2009)
  [arXiv:0808.3498 [hep-th]].
  %%CITATION = PHRVA,D79,021901;%%
}

\Title{\vbox{\baselineskip12pt
}}
{\vbox{\centerline{Holographic Lovelock Gravities and  Black Holes}
\vskip.06in
}}
\centerline{Jan de Boer${}^a$, Manuela Kulaxizi${}^a$, and Andrei Parnachev${}^b$}
\bigskip
\centerline{{\it ${}^a$Department of Physics, University of Amsterdam }}
\centerline{{\it Valckenierstraat 65, 1018XE Amsterdam, The Netherlands }}
\centerline{{\it ${}^b$C.N.Yang Institute for Theoretical Physics, Stony Brook University}}
\centerline{{\it Stony Brook, NY 11794-3840, USA}}
\vskip.1in \vskip.1in \centerline{\bf Abstract}
\noindent
We study holographic implications of Lovelock gravities in AdS
spacetimes.
For a generic Lovelock gravity in arbitrary spacetime dimensions
we formulate the existence condition of asymptotically AdS
black holes.
We consider  small fluctuations around these black holes and determine
the constraint on Lovelock parameters by demanding causality of the
boundary theory.
For the case of cubic Lovelock gravity in seven
spacetime dimensions we compute the holographic Weyl
anomaly and determine the three point functions of
the stress energy tensor in the boundary CFT.
Remarkably, these correlators happen to
satisfy the same relation as the one imposed by supersymmetry.
We then compute the energy flux; requiring it to be positive
is shown to be completely equivalent to requiring causality
of the finite temperature CFT dual to the black hole.
These constraints are not stringent enough to place any positive lower bound
on the value of viscosity.
Finally, we conjecture an expression for the energy flux valid
for any Lovelock theory in arbitrary dimensions.

\vfill

\Date{December 2009}

%\draftmode

\newsec{Introduction and summary}

\noindent
Lovelock gravity \Lovelock\ is the most general classical theory of gravity whose
equations of motion contain at most second derivatives of the metric.
The simplest example of Lovelock gravity is just the usual Einstein-Hilbert
gravity.
As the number of spacetime dimensions grows, Lovelock gravity allows more
and more higher derivative terms.
For example, in five dimensions one can add a term quadratic in the Riemann
tensor, which gives rise to Gauss-Bonnet gravity.
Important features of Lovelock gravity are the absence of ghosts
in Minkowski backgrounds \refs{\ZwiebachUQ-\ZuminoDP} and  the equivalence
between metric and Palatini formulations \ExirifardJabbari.
One may wonder what role these higher derivative gravity theories
play in the AdS/CFT correspondence \refs{\Maldacena\Witten-\GKP}.
The first interesting nontrivial example is Gauss-Bonnet gravity in
an $AdS_5$ background which defines a four dimensional CFT.
One can compute Weyl anomalies for this theory and find that
the ratio of the two central charges, $a$ and $c$, depends nontrivially
on the value of the Gauss-Bonnet coupling $\lambda_1$, and goes to one
in the limit of vanishing $\lambda_1$ which corresponds to Einstein-Hilbert
gravity  \refs{\NojiriMH\BriganteLMSYa-\BriganteLMSYb}.
Plenty of CFTs with $a\neq c$ are known, and having a gravity dual
of such CFTs would be desirable.

In addition to the special features of
Lovelock theories mentioned above, Gauss-Bonnet theory has been shown
to have a peculiar property in the context of the AdS/CFT correspondence.
Consider an $AdS_5$ solution of Gauss-Bonnet gravity which defines a dual four-dimensional CFT.
Requiring positivity of the energy flux in a four-dimensional CFT places certain constraints on the
values of the two parameters which determine the angular distribution of the
energy flux, $t_2$ and $t_4$ \HofmanMaldacena.
Assuming supersymmetry, these constraints can be reformulated as bounds on
the ratio of $a$ and $c$ central charges of the CFT, which
in turn imply bounds on the value of Gauss-Bonnet coupling, $\lambda_1$.
Exactly the same bounds on $\lambda_1$ have been obtained by requiring causality
of the finite temperature CFT \refs{\BriganteLMSYa\BriganteLMSYb\BuchelMyers-\Hofman}.
The story repeats itself for the six-dimensional CFT dual to Gauss-Bonnet gravity
in $AdS_7$ \deBoerPN.
In Ref. \deBoerPN\ we computed the values of $t_2$ and $t_4$
in terms of three independent coefficients which determine
the three and two point functions of the stress energy tensor.
We observed that $t_4$ is proportional to the combination of
the three point functions which vanishes in the supersymmetric theory.
We computed the holographic Weyl anomaly for Gauss-Bonnet gravity
and found that $t_4=0$ while the positivity of energy flux condition
constrains the value of the Gauss-Bonnet coupling $\lambda_1$ to lie
within a certain interval.
We have also analyzed the propagation of a graviton with helicity 2
in the black hole background and found that the bound on $\lambda_1$
coming from causality precisely matches the lower bound on $\lambda_1$
from flux positivity.
Generalizing the matching between causality and positivity of the energy flux
to other polarizations in GB theories of arbitrary dimensions
was achieved in \refs{\CamanhoVW-\BuchelSK}.
In particular, \BuchelSK\ computed the three point functions of
the stress-energy tensor and determined $t_2$ and $t_4$ in the CFTs dual to
GB gravities in any dimensions.

In this paper we study Lovelock theories with negative cosmological
constant, paying special attention to
cubic Lovelock gravity in an $AdS_7$ background.
In this case there are two independent coefficients which multiply the  Gauss-Bonnet term
and the term cubic in the Riemann tensor.
In general the $\OO(R^3)$ term is expected to describe a generic
non-supersymmetric case.
However the third order Lovelock term does not contribute to the
three-point functions of gravitons in flat space \MetsaevTseytlin.
Yet, we show that the corresponding contribution in the $AdS$ background does
not vanish and hence the positivity of energy flux condition results in
nontrivial constraints on both the Gauss-Bonnet and third order Lovelock
coefficients, $\lambda_1$ and $\lambda_2$.

To find the constraints imposed by causality of the boundary theory
we analyze the space of black hole solutions.
We are able to formulate the conditions of black hole existence and
compute the causality constraint for the graviton of helicity 2 in Lovelock
gravity of arbitrary dimensionality.
In the case of cubic Lovelock in seven spacetime dimensions
we write down the black hole solutions explicitly; interestingly, for some
values of the Lovelock parameters there are AdS solutions but no asymptotically
AdS black holes.
Remarkably, we find that this solution structure ensures
that the causality constraint coincides with the energy positivity constraint.
Another remarkable feature of Lovelock gravity is vanishing $t_4$,
one of the parameters that determine the angular distribution of the energy flux.
As explained in  \deBoerPN, $t_4$ is proportional to the linear combination of
parameters which vanishes when the minimal supersymmetry is assumed.
Constraints from flux positivity, together with the condition
for the black hole to exist, restrict the value of the Gauss-Bonnet coupling from
above.
Interestingly, this restriction is not sufficient to
place a positive lower bound on the viscosity/entropy ratio.

The rest of the paper is organized as follows.
In the next Section we recall the description of the
most generic Lovelock gravity in any spacetime dimensions
and formulate the conditions for the asymptotically AdS back holes to exist.
We also analyze the propagation of gravitons of helicity 2 in these black hole backgrounds
and derive bounds on the Lovelock coefficients resulting from causality.
In Section 3 we apply these results to the case of cubic Lovelock theory
in seven spacetime dimensions.
In particular, we write down explicit formulas for the black hole
solutions.
To the best of our knowledge, some of these solutions have not appeared
in the literature before.
In Section 4 we compute the holographic Weyl anomaly in cubic Lovelock
gravity in seven dimensions and determine the bounds on the Lovelock coefficients
from flux positivity.
We find that there is precise matching between the causality and
positivity of flux bounds, just like in the Gauss-Bonnet case.
We discuss our results in Section 5 where, among other things, we discuss
implications of our results to the viscosity/entropy ratio and  make
an educated guess for the  energy flux in any Lovelock
theory in arbitrary spacetime dimensions.

\newsec{Lovelock gravity, black holes and fluctuations}

\noindent Lovelock gravity \Lovelock\ is the most general classical theory of gravity which
yields covariant second order field equations, i.e., the equations of motion
which contain only up to second order derivatives of the metric tensor. The Lovelock
action for a $d+1$-dimensional spacetime is
\eqn\LovelockAction{S= \int d^{{d+1}}x \sqrt{-g}\sum_{p=0}^{[{d\over 2}]} \beta_{p} {\cal{L}}_{p}}
where $[{d\over 2}]$ denotes the integral part of ${d\over 2}$, $\beta_p$
is the $p$-th order Lovelock coefficient and ${\cal{L}}_p$ defined as
\eqn\Lps{{\cal{L}}_p={1\over 2^{p}} \delta^{\mu_1\nu_1\cdots \mu_p\nu_p}_{\rho_1\sigma_1\cdots\rho_p\sigma_p }
R^{\rho_1\sigma_1}_{\qquad\mu_1\nu_1}\cdots R^{\rho_p\sigma_p}_{\qquad\mu_p\nu_p} }
is the Euler density of a $2p$--dimensional manifold.
Here $\delta^{\mu_1\nu_1\cdots \mu_p\nu_p}_{\rho_1\sigma_1\cdots\rho_p\sigma_p }$ denotes the totally antisymmetric
product of Kronecker delta symbols while $R^{\rho_q\sigma_q}_{\qquad\mu_q\nu_q}$ is the Riemann curvature tensor.
By construction it is clear that in $d+1$ dimensions all Lovelock ${\cal{L}}_p$ terms for which $p\geq [{d\over 2}]$
either vanish (for $p>d/2$) or are total derivatives (for $p=d/2$)  and do not contribute to the equations of motions.
The $p=0$ and $p=1$ terms correspond to the cosmological constant and Ricci scalar respectively.
Our conventions are such that $L=1$ while $\beta_0=d(d-1)$ and $\beta_1=1$.

To determine the black brane solutions of Lovelock gravity
we consider the metric
\eqn\metric{ds^2 = -f(r) dt^2 + \frac{dr^2}{f(r)} + r^2 \sum_{i=1}^{d-1}
dx_i^2 .}
%Let us consider $p$-th order Lovelock in $d+1$ dimensions with Lagrangian
%\eqn\Ltotal{{\cal L}=\sum_{p=0}^{p} \lambda_p {\cal{L}}_p}
For this ansatz, the only non-vanishing components of the Riemann
tensor (up to symmetries) are
\eqn\riemann{R^{tr}{}_{tr} = -f''/2,\quad R^{ti}{}_{ti} = R^{ri}{}_{ri} =
-\frac{f'}{2r}, \quad R^{ij}{}_{ij} = -\frac{f}{r^2}.}
where primes indicate differentiation with respect to the variable $r$.
This makes it relatively straightforward to evaluate the Lovelock
action on a black hole solution of the form \metric.
The $p$-th Lovelock term evaluates as
\eqn\pthl{\int d^{d+1}x\sqrt{-g}{\cal L}_p = (2 p)!! \int dr\,dt\,dx^{d-1}\, r^{d-1}(Q_1+Q_2+Q_3+Q_4)}
where
\eqn\Qs{\eqalign{ Q_1 & = -\frac{f''}{2} C^{d-1}_{p-1} \left( \frac{-f}{r^2}
\right)^{p-1} \cr
Q_2 & = (d-1)(d-2) \left( \frac{-f'}{2r}  \right)^2
C^{d-3}_{p-2} \left( \frac{-f}{r^2} \right)^{p-2} \cr
Q_3 & = -\frac{f'}{r} (d-1) C^{d-2}_{p-1} \left( \frac{-f}{r^2}
\right)^{p-1} \cr
Q_4 & = C^{d-1}_{p} \left( \frac{-f}{r^2} \right)^{p}
}}
and $C^d_p$ is the number of ways you can pick $p$ pairs from
$d$ numbers, i.e. $C^d_p=d!/((d-2p)! (2p)!!)$.
After some algebra, we find
\eqn\pthLs{\int d^{d+1}x\sqrt{-g}{\cal L}_p = (-1)^p\frac{(d-1)!}{(d-2p+1)!} \int dr\,dt\,dx^{d-1}\,
\partial_r^2 \left[ r^{d-2p+1} f^p \right].}

This is a total derivative, so varying $f$ does not lead to a useful
equation of motion. To find the equations of motion, we have to vary
the metric and curvature tensor with respect to arbitrary variations of the metric.
The variation of the Riemann curvature tensor reads,
denoting $\zeta_{\mu}{}^{\nu}= \frac{1}{2} g_{\mu\alpha} \delta
g^{\alpha\nu}$,
\eqn\varRR{
\delta R_{\alpha\beta}{}^{\gamma\delta} = \frac{1}{2} \{
\nabla_{[\alpha}, \nabla^{[\gamma} \}\zeta_{\beta]}{}^{\delta]}
-\frac{1}{2} \zeta_{[\alpha}{}^{\epsilon}
R_{\beta]\epsilon}{}^{\gamma\delta} - \frac{1}{2}
\zeta_{\epsilon}{}^{[\gamma} R_{\alpha\beta}{}^{\delta]\epsilon} .}
The first term doesn't yield any contribution, because we can
partially integrate the covariant derivative in the Lovelock
action and because all indices are totally antisymmetrized, the
result will vanish because of the Bianchi identity.
The second term will contribute and one
finds for the metric \metric\ that $\delta R_{\mu\nu}{}^{\mu\nu}=
(\zeta_{\mu}{}^{\mu} + \zeta_{\nu}{}^{\nu}) R_{\mu\nu}{}^{\mu\nu}$
where one does not sum over $\mu$ and $\nu$.

To get the equation of motion, we should in principle
work out the variation of the action with respect to $\zeta_{\mu}{}^{\nu}$.
One finds that the variations with respect to $\mu\neq\nu$ vanish identically,
the variations with respect to $\zeta_r{}^r$ and $\zeta_t{}^t$ yield a term
proportional to $Q_3+2Q_4$, and the variation with respect to $\zeta^i{}_i$ yields a term
proportional to a linear combination of $Q_3+2Q_4$ and $Q_1+Q_2+Q_3+Q_4$.
By putting the terms proportional to $Q_3+2Q_4$ equal to zero, we obtain the equations of motion for $p$-th order
Lovelock gravity in $d+1$ dimensions
\eqn\eqomh{\left[ \sum_p \beta_p (-1)^p{(d-1)!\over (d-2p)!} r^{d-2p} f^p \right]'=0.}
The other equation that one obtains, which is proportional to $Q_1+Q_2+Q_3+Q_4$,
is a linear combinations of this field equation and its $r$-derivative,
and therefore contains no additional information.

It is convenient to define $\hat{\lambda}_p$ so that
\eqn\lambdatil{(d-1)\hat{\lambda}_p=\beta_p (-1)^p{(d-1)!\over (d-2p)!}}
With our conventions for $\beta_0$ and $\beta_1$ we have that $\lambdah_0=1$ and $\lambdah_1=-1$.
If we denote
\eqn\PQdef{R(r)= \sum_p \lambdah_p r^{d-2p} f^p \qquad\qquad Q(r)= {\p R\over\p f} =\sum_p p \lambdah_p r^{d-2p}
f^{p-1}.}
we can express the equation of motion \eqomh\ as
\eqn\eqomtwo{R(r)=K\Rightarrow \sum_p \lambdah_p \left(\frac{f}{r^2}\right)^p = K/r^d}
for some constant $K$. Evaluating $R(r)=K$ at the horizon we find that $\lambdah_0=K/r_+^d$,
which leads to $K=r_{+}^d$.
% we see that $K$ needs to be positive, and
%since $\beta$ was negative to have a positive mass black hole (at
%least that seems to be the right criterion) it follows that
%$\sum_p p \lambda_p \alpha^{p-1}<0$. Therefore the bound becomes
% \be \sum_p p ((d-2)(d-3) + 2 d (p-1) ) \lambda_p
%\alpha^{p-1} \leq 0 . \label{eq12}
%\ee
Since \metric\ describes an AdS black hole solution we expect $f(r)$ to behave for large $r$ as
\eqn\fexp{f(r)= \alpha r^2 + \gamma r^{2-d} + \ldots}
Combining \eqomtwo\ and \fexp\ we deduce that
\eqn\alphaeq{\sum_p \lambdah_p \alpha^p = 0}
and
\eqn\betadef{\gamma={r_{+}^d\over \sum_p p\lambdah_p \alpha^{p-1}}}
Obviously $\gamma$ is related to the black hole mass, so we need to assume that
$\gamma$ is negative. Note that since $r_{+}$ is positive, this leads to
\eqn\den{\sum_p p\lambdah_p \alpha^{p-1}<0}
%This inequality will prove crucial for the results of section ...

Interestingly, \eqomtwo\  implies that $f/r^2$ is monotonic.
There cannot be two values of $r$ for which $f/r^2$ takes
identical values, because then $r_{+}^d/r^d$ would have to take
identical values as well, which implies $r_1=r_2$.
Thus, most real solutions of \eqomtwo\ will give rise to
acceptable black hole solutions. In fact, as long as $\gamma<0$ and
obviously $\alpha>0$, $f/r^2$ increases for very large $r$, and
therefore, $f/r^2$ has to increase everywhere. Since $r_{+}^d/r^d$ goes
to infinity as $r\rightarrow 0$, $f/r^2$ cannot remain bounded as
$r\rightarrow 0$. Since it starts of at the value $\alpha$ at
$r=\infty$, decreases as we decrease $r$ and cannot remain
bounded, it must become zero somewhere and there will be a
horizon. So as long as $f/r^2$ remains real and $\gamma<0$, this is indeed a
black hole type solution. The only subtlety is that a real
solution of $f$ might cease to exist at some finite value of $r$.

So defining
\eqn\pdef{ P(x)=\sum_p \lambdah_p x^p ,}
for a proper black hole solution, $P(x)$ must be monotonously
decreasing between $0$ and $\alpha$ with $\alpha$ the smallest positive root of
$P(x)$. To determine therefore the conditions for a black hole solution to exist
we should require that the extrema of $P(x)$, if any, occur outside the region $x\in[0,\alpha]$.
In the following section we will carefully analyze these requirements in the context of 3rd
order Lovelock gravity.
%We will determine the precise conditions the Lovelock parameters should satisfy for
%a black hole solution to exist.

Let us now move on to the study of metric fluctuations. We will restrict ourselves to fluctuations around the black hole solution \metric\
in the scalar channel, \ie, $\phi = h_{12}$. Other graviton polarizations can be studied in a similar fashion.
The form of the perturbed metric is
\eqn\perds{ds^2=- f(r)dt^2+{dr^2\over f(r)}+r^2 \left[\sum_{i=1}^{d-1}dx_i^2+2 \phi(t,r,x_{d-1})dx_1dx_2\right]}
Since $\phi$ only depends on the $(t,r,x_{d-1})$ directions of spacetime its Fourier transform can be written as
\eqn\Fourier{\phi(t,r, x_{d-1})=\int {d\omega d q \over (2\pi)^2} \varphi(r) e^{-i\omega t+i q x_{d-1}} \qquad k=(\omega,0,0,\cdots ,0,q)}
The equations of motions for $\varphi$ can be found by expanding the Lagrangian \Lps\ to second order in the fluctuating field.
Alternatively one can substitute the metric \perds\ in the equations of motion and expand to linear order in $\varphi$. The result is
\eqn\eqomfl{T_2 \varphi''(r)+T_2' \varphi'(r)+T_0 \varphi(r)=0}
where primes indicate differentiation with respect to the variable r and $T_2(r),T_0(r)$ are expressed
with the help of \PQdef\ as follows
\eqn\Tidef{T_2={d-3 \over 2} r^2 f(r) \left(\p_r Q\right), \qquad\qquad T_0={T_2(r)\over a^2f(r)^2}\omega^2-{1\over 2} f(r) \left(\p_r^2 Q\right) q^2}
%with
%\eqn\bwbq{b_{\omega}={T_2(r)\over f(r)^2} \qquad\qquad b_{q}=-{1\over 2} f(r) a^2 \left(\p_r^2 Q\right) }
%\eqn\Tidef{\eqalign{T_0&=\tom^2 \left[3r^3\left(-2 r+\lambda_1 L^2 f'(r)\right)+3r f(r)\left(2\lambda_1 L^2 r-\lambda_2 L^4 f'(r)\right)\right]+\cr
%&+\tq^2 a^2 L^2 f(r)\left[-4\lambda_1 L^2 r f'(r)+\lambda_2 L^4 f'(r)^2+r^2\left(6-\lambda_1 L^2 f''(r)\right)-f(r)\left(2\lambda_1 L^2-\lambda_2 L^4 f''(r)\right)\right]\cr
%T_1&=3 a^2 L^4 f(r) \left[r^3 f'(r)\left(-2 r+\lambda_1 L^2 f'(r)\right)
%-L^2 f(r)^2 \left(-6r\lambda_1 +\lambda_1 L^2 \left(2 f'(r)+r f''(r)\right)\right)\right.\cr
%&\left. +r f(r)\left(8\lambda_1 L^2 r f'(r)-2 \lambda_2 L^4 f'(r)^2+r^2\left(-10+\lambda_1 L^2 f''(r)\right)\right)\right]\cr
%T_2&=3 a^2 L^4 r f(r)^2\left[r^2\left(-2 r+\lambda_1 L^2 f'(r)\right)+f(r)\left(2\lambda_1 L^2 r-\lambda_2 L^4 f'(r)\right)\right] }}
%Primes indicate differentiation with respect to the variable r and we defined $\tom=\omega L^2$
%and $\tq=q L^2$.
Note that we rescaled the time coordinate $t\rightarrow a t$ so as to set the boundary speed of light to unity.
One easily checks that if $f(r)$ is of the form given in \fexp\ then $\alpha={1\over a^2}$.

It is convenient to make a coordinate transformation from $r$ to $y$ according to $a f(r)\p_r y(r)=1$
and place equation \eqomfl\ in Schrodinger form
\eqn\eqomSchr{-\p_{y}^2\Psi+\left[q^2 c_g^2(y)+V_1(y)\right]\Psi=\omega^2 \Psi}
Here $\Psi(y)$ is defined as $\Psi=\varphi {T_2^2\over \sqrt{f}}$ while
\eqn\cgsdef{\eqalign{c_g^2={a^2 \over d-3}{ f\over r}&{\p_r^2 Q\over \p_r Q} \qquad\qquad
V_1(y)=-a^2 f^2 h(y)-{1\over a}\sqrt{f} \p_y \left[f^{-2} \p_y \sqrt{f}\right]\cr
& h(y)=-{1\over 2 a^2 f} \p_y\left[{1\over f} {\p_y T_2\over T_2}\right]-{1\over 4 a^2 f^2}\left[{\p_y T_2\over T_2}\right]^2} . }

%To do this, we follow two steps:
%We first define a new function $\Phi(r)$ through
%\eqn\sone{\ln{\Phi}=\ln{\varphi}+{1\over 2}\int {T_1\over T_2}}
%which brings eq. \eqomfl\ into the standard form
%\eqn\eqomflsf{\Phi''(r)+W(r)\Phi(r)=0 \qquad W(r)={T_0\over T_2}-{1\over 2}\left[{T_1'T_2-T_2'T_1 \over T_2^2}\right]-{1\over 4}\left({T_1\over T_2}\right)^2 }
%From eq. \Tidef\ observe that we can write $T_0(r)$ as $T_0(r)=b_{\omega}(r)\tom^2+b_{q}(r)\tq^2$
%where $b_{\omega}(r),b_{q}(r)$ are functions of the radial coordinate $r$ alone.
%This implies that $W(r)$ can be expressed as
%\eqn\Whdef{W(r)={b_{\omega}\over T_2}\tom^2+{b_{q}\over T_2}\tq^2+h(r) \qquad h(r)=-{1\over 2}\left[{T_1'T_2-T_2'T_1 \over T_2^2}\right]-{1\over 4}\left({T_1\over T_2}\right)^2}
%Substitute then $\Phi(r)$ with $\Psi(r)$ defined as $\Psi(r)=\left({b_{\omega}\over T_2}\right)^{{1\over 4}} \Phi(r)$
%and subsequently make a coordinate transformation from $r$ to $y$ according to:
%\eqn\tvry{\p_r y(r)=\sqrt{{b_{\omega}\over T_2}}}
%Eq. \eqom\ is finally expressed as:
%\eqn\eqomSchr{-\p_{y}^2\Psi+\left[\tq^2 c_g^2(y)+V_1(y)\right]\Psi=\tom^2 \Psi}
%where
%\eqn\cgsdef{c_g^2=-{ b_{q} \over b_{\omega} } \qquad\qquad
%V(y)=-{T_2\over b_{\omega}}h(y)-\left({b_{\omega}\over T_2}\right)^{-{1\over 4}}
%\p_y \left[ {b_{\omega}\over T_2} \p_y \left( {b_{\omega}\over T_2} \right)^{-{1\over 4}}\right] }

We are now ready to study the full graviton wave function \eqomSchr\ .
Note that $y(r)$ is a monotonically increasing function of $r$ with $y\ra 0$ at the boundary $r>>r_{+}$ and $y\ra -\infty$
at the horizon $r=r_{+}$.
%One can show that $V_1(y)$ is also a monotonic function of $y$ which blows up as $y^{-2}$ for $y\ra 0$.
Following \refs{\BriganteLMSYa,\BriganteLMSYb} we consider \eqomSchr\ in the limit $q\ra\infty$.
In this case, $q^2 c_g^2(y)$ provides the dominant contribution to the potential except for a small region $y> -{1\over q}$.
It is therefore reasonable to approximate the potential with $c_g^2(y)$ for all $y<0$ and replace it with an infinite wall
at $y=0$.
Consider now the behaviour of $c_g^2(y)$ in the proximity of the boundary $y=0$. This is easier to
analyze in the original variable $r$. In particular,
\eqn\cgsex{c_g^2=1+C{r_{+}^d\over r^d}+\cdots}
with $C$ a function of the Lovelock coefficients $\lambda_p$ equal to
\eqn\Cexpd{C={1\over \alpha\left(d-2\right)\left(d-3\right) } {\sum_p p ((d-2)(d-3) + 2 d (p-1) ) \lambdah_p
\alpha^{p-1}\over \left(\sum_p p\lambdah_p \alpha^{p-1}\right)^2 }}
To arrive at \Cexpd\ we used eqs \cgsdef\ , \betadef\ as well as $\alpha={1\over a^2}$.
%It will be convenient for the analysis of the next section to have an expression for $C$ in $d=6$ dimensions
%as a function of the $AdS$ radius and $\lambda_1$ defined in \lambdapdef\
%\eqn\cgexp{C=-a^4 {5\lambda_1+a^2 \left(-8+9 a^2\right)\over \left[\lambda_1+a^2\left(-2+3 a^2\right)\right]^2} }

Observe that when $C$ is positive, $c_g^2(r)$ attains a maximum value (greater than one) within the bulk geometry.
The same is of course true for $c_g^2$ as a function of the $y$ coordinate.
However, the existence of a maximum for $c_g^2(y)$ implies the existence of metastable states from the
point of view of the boundary theory. Moreover, the group velocity of these states can be determined with the
WKB approximation to be greater than unity, i.e., the group velocity approaches $c_{g,max}>1$ at the same time
the phase velocity ${\omega\over q}$ tends to $c_{g,max}$ \refs{\BriganteLMSYa,\BriganteLMSYb}.
 Hence, for values of the Lovelock parameters $\lambda_p$ such that $C>0$,
the boundary theory violates causality \refs{\BriganteLMSYa,\BriganteLMSYb}.
 Gauge-gravity duality is therefore applicable only when $C$ is negative.
From \Cexpd\ this leads to the condition
\eqn\causalitycondition{\sum_p p ((d-2)(d-3) + 2 d (p-1) ) \lambdah_p
\alpha^{p-1}<0}
for any $p$-th order Lovelock theory of gravity in $d+1$ dimensions.

\newsec{Black holes of 3rd order Lovelock gravity}

\noindent In this section we will analyze asymptotically AdS black hole solutions
of 3rd order Lovelock gravity in seven dimensions with flat horizon.
In Section 3.1. we will apply the results of Section 2
to determine the parameter space of these black hole solutions.
In Section 3.2. we write down the solutions explicitly.
For completeness we write here
the action of third order Lovelock gravity in $6+1$ dimensions
\eqn\Lthree{S= \int d^{7}x \sqrt{-g}\left[{30\over L^2}+R+{\lambda_1 L^2\over 12}{\cal L}_2+{\lambda_2 L^4\over 72}{\cal L}_3\right].}
Note that $\lambda_1$ and $\lambda_2$ are related to the Lovelock parameters $\hat\lambda$
defined in \lambdatil\ as
\eqn\lambdapdef{ \lambda_1=\lambdah_2, \qquad \lambda_2=-3 \lambdah_3  }

\subsec{Existence of black hole solutions}

\noindent We start by noting that the action \Lthree\ admits AdS spacetime as a solution.
In the $d=6$ case
the AdS radius is related to the parameter $L$ in \Lthree\ via
\eqn\defarem{   L_{AdS}=a  L  }
where $a$ satisfies
\eqn\lads{  \lambda_2=3 a^2 \left[ \lambda_1-a^2+
          a^4 \right]   }
It is clear that AdS solutions exist for all $\lambda_1$ and
$a^2>0$, since the value of $\lambda_2$ can be determined from \lads.
It is not hard to map this parameter space to the $(\lambda_1,\lambda_2)$ plane.
However it is more convenient to parameterize  the
solutions using the variables $\lambda_1$ and $\alpha=1/a^2$.
Consider a black hole
which asymptotes to $AdS_7$  with the radius $a=1/\sqrt{\alpha}$.
As explained in Section 2 such a black hole exists when
$P(x)$ defined by \pdef\ is a monotonically decreasing function of $x$ between $x=0$ (where $P(0)=1$)
and $x=\alpha$, where $\alpha$ is the smallest root of $P(x)$.
Using $\lambdah_0=1,\,\lambdah_1=-1,\,\lambdah_2=\lambda_1$ and $\lambdah_3=-{\lambda_2\over 3}$,
we can write
\eqn\pdef{  P(x)=1-x+\lambda_1 x^2-{\lambda_2\over3} x^3  }
Let us now fix the value of $\lambda_1$ and find a constraint on
$\alpha$ which follows from this condition.
Suppose $\lambda_1>1/4$.
%For $\lambda_2>\lambda_1^2$ the existence of black hole is guaranteed,
%since $P(x)$ is a monotonic function of $x$.
For  $\lambda_2<\lambda_1^2$, $P(x)$ develops a minimum
at
\eqn\xmdef{  x_-={\lambda_1-\sqrt{\lambda_1^2-\lambda_2}\over\lambda_2}  }
and a maximum at
\eqn\xpdef{  x_+={\lambda_1+\sqrt{\lambda_1^2-\lambda_2}\over\lambda_2}  }
The black hole solution exists as long as $P(x_-)\leq0$.
Hence, $\alpha$ is constrained from above by the value of $x=\alpha_{max}$ such
that
\eqn\alphac{   P(\alpha_{max})=0,\qquad  {\p P\over \p x} (\alpha_{max})=0  }
simultaneously.
This gives two solutions; the upper  solution corresponds to $P(x_+)=0$,
while the lower one is what we need,
\eqn\alphacs{  \alpha_{max}={1-\sqrt{1-3\lambda_1}\over\lambda_1}    }
Something interesting happens at $\lambda_1=1/3$ where $\alpha_{max}=3$.
For $\lambda_1>1/3$ there is no real solution of eq. \alphac.
In this region the existence of the black hole solution is equivalent to
the requirement that $P(x)$ is a monotonic function, i.e. $\lambda_2>\lambda_1^2$.
With the help of \alphaeq\ this can be restated as
\eqn\condlone{  {3\alpha^2-\sqrt{3 (4-\alpha)\alpha^3}\over2\alpha^3}\leq\lambda_1\leq
               {3\alpha^2+\sqrt{3 (4-\alpha)\alpha^3}\over2\alpha^3},\quad \lambda_1>{1\over3}   }

The analysis above can be repeated  for $0\leq\lambda_1\leq 1/4$.
Now the value of $\alpha$ is again bounded from above by $\alpha_{max}$
which is a solution of \alphac.
The upper solution with positive $\lambda_2$ again corresponds to
$P(x_+)=0$, while the lower solution \alphacs, now with negative $\lambda_2$,
is again the upper bound on $\alpha$.
The cases of $\lambda_1=0$ and $\lambda_1\leq0$ can be analyzed in a
similar manner.
The result is that black hole solutions exist whenever one of the two conditions,
either \condlone, or
\eqn\bhexist{  0<\alpha<{1-\sqrt{1-3\lambda_1}\over\lambda_1} , \qquad \lambda_1<{1\over3}   }
is satisfied.
Note that the curves defined by \condlone\ and \bhexist\ meet at the point $(\lambda_1=1/3,\alpha=3)$.
\midinsert\bigskip{\vbox{{\epsfxsize=3in
        \nobreak
    \centerline{\epsfbox{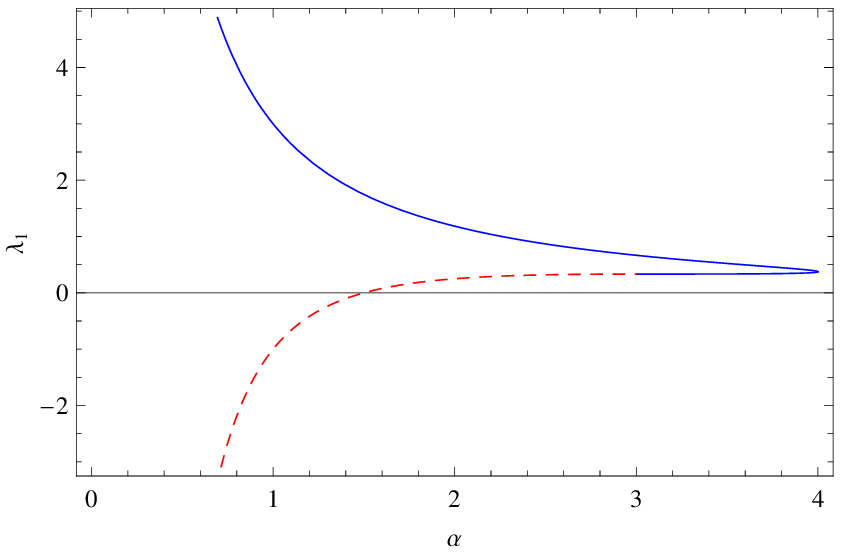}}
        \nobreak\bigskip
    {\raggedright\it \vbox{
{\bf Fig 1.}
{\it  Black holes exist in the region $\alpha\leq \alpha(\lambda_1)$ bounded from the right by the  curve.
Red dashed part of the  curve corresponds to eq. \condlone. Blue solid part of the  curve is determined by \bhexist.
AdS solutions exist for all $\lambda_1$ and $\alpha>0$.
}}}}}}
\bigskip\endinsert
\noindent
%
%Consider now the requirement that the denominator in ... is positive.
%One can show that for positive $a^2$, this condition is precisely \bhexist.
%Hence for the third order Lovelock gravity the causality constraint precisely matches the positivity of
%energy constraint.
%The non-trivial parameter space of black hole solutions plays a crucial
%role in this matching.
%
The results of this discussion are summarized in Fig. 1.
An AdS solution exists for any values of $\lambda_1$ and any positive $\alpha$.
The region of the parameter space where black holes exist is
bounded by the requirement that $\alpha$ is smaller than the value
determined by the curve in Fig. 1.

\subsec{Explicit Solutions}

\noindent Explicit black hole solutions for Lovelock gravity have been discussed in many places in the literature.
A non exhaustive list of references is \refs{\MyersSimon\Wheeler\DehghaniPourhasan\DehghaniAH\DehghaniBH\Takahashia\Takahashib\Caia-\Caib}.
To our knowledge however, it has always been assumed that only the real
roots of eq. \eqomtwo\ describe consistent black holes. Here we will explore other possibilities and find that they
indeed satisfy the criteria for valid black hole solutions as described in the previous subsection. In fact,
the analysis here can be exactly matched to the previous one although it is the result of independent reasoning.

Following \DehghaniBH\ we write the black hole metric in the form
\eqn\bhsol{\eqalign{ds^2&=-a^2 f(r)dt^2+{dr^2\over f(r)}+{r^2\over L^2}\sum_{i=1}^{5}dx_i^2\cr
f(r)&={r^2\over L^2} {\lambda_1\over\lambda_2}X(r)\cr
X(r)&=1+\left(J(r)+ \sqrt{J(r)^2+G^3}\right)^{{1\over 3}}+\left(J(r)-\sqrt{J(r)^2+G^3}\right)^{{1\over 3}}
}}
where $J(r),\, G,\, a$ are defined as
\eqn\JGsol{\eqalign{G&={\lambda_2\over\lambda_1^2}-1\cr
J(r)&=1-{3\over 2}{\lambda_2\over\lambda_1^2}+{3\over 2}{\lambda_2^2\over\lambda_1^3}\left(1-{r_{+}^6\over r^6}\right)\equiv
J_{\infty}-\left({1\over 2}+J_{\infty}+{3\over 2}G\right){r_{+}^6\over r^6}\cr
a^2&=\left[\lim_{r\ra\infty}{ {L^2\over r^2}f(r)}\right]^{-1}={\lambda_2\over\lambda_1} X_{\infty}^{-{1}} }}
and the AdS radius
\eqn\AdSRadius{L_{AdS}=a L}
with $L$ related to the cosmological constant.
%\left( 1+\left(\sqrt{J_{\infty}^2+G}+J_{\infty}\right)^{{1\over 3}}-{G\over\left(\sqrt{J_{\infty}^2+G}-J_{\infty}\right)^{{1\over 3}}} \right)^{-1}
Notice that $X(r)$ in $L=1$ units is related to the variable $x$ of the previous subsection through
$X={\lambda_1\over\lambda_2}x$. For our purposes it is useful to express $X(r)$ as
\eqn\Qtwo{X(r)=\left\{\eqalign{
1+\left(J(r)+ \sqrt{J(r)^2+G^3}\right)^{{1\over 3}}-{G\over\left(J(r)+\sqrt{J(r)^2+G^3}\right)^{{1\over 3}}} \qquad \lambda_1> 0 \cr
1+\left(J(r)- \sqrt{J(r)^2+G^3}\right)^{{1\over 3}}-{G\over\left(J(r)-\sqrt{J(r)^2+G^3}\right)^{{1\over 3}}} \qquad \lambda_1<0
       }\right. }
With $X(r)$ written in this form we see that the solution is determined by the cubic root of the function
$M_{\pm}(r)\equiv\pm\sqrt{J(r)^2+G^3}+J(r)$, where the $\pm$ sign matches the sign of $\lambda_1$.
This means that the equations of motion admit {\it{three}} different solutions, classified
by the cubic roots of $M_{\pm}(r)$ left unspecified in \bhsol (and \Qtwo ). This is also clear from the
form of \eqomtwo.

Here we examine which of the three (if not all) and under which conditions, constitute
a black hole solution. We will mostly use $X(r)$ as expressed in \Qtwo\ but refer to the one in \bhsol\ when convenient.
Different cases depend on the sign of $\Delta(r)\equiv J^2(r)+G^3$ sitting under the square root in \bhsol.
It is useful to consider them separately.

\item {(1)} Case I:$\quad G>0\Leftrightarrow \lambda_2>\lambda_1^2$

\noindent Here $\Delta(r)\equiv J^2(r)+G^3$ is positive for all $r$ which implies $M_{\pm}(r)=\pm\sqrt{J^2(r)+G^3}+J(r)$ is real.
At the same time $M_{+}(r)$ is positive for all $r$ whereas $M_{-}(r)$ is negative.
The three cubic roots of $M_{\pm}(r)$ are $|M_{\pm}(r)|^{{1\over 3}} e^{i {\chi_{\pm}+2n\pi\over 3}}$ with $n=0,1,2$.
Naturally, $\chi_{+}=0$ for $M_{+}$ and $\chi_{-}=\pi$ for $M_{-}$.
%and $M(r)^{{1\over 3}}$ the real and positive cubic root of $M(r)$.
Notice that $X(r)$ for all three solutions $(n=0,1,2)$ can be written as
\eqn\Qone{X(r)=\left\{\eqalign{
&1+|M_{+}(r)|^{{1\over 3}} e^{i {2n\pi\over 3}}-{G\over|M_{+}(r)|^{{1\over 3}}}e^{-i {2n\pi\over 3}}\qquad\qquad \lambda_1\geq 0\cr
&1+|M_{-}(r)|^{{1\over 3}} e^{i {\pi+2n\pi\over 3}}-{G\over|M_{-}(r)|^{{1\over 3}}}e^{-i {\pi+2n\pi\over 3}}\qquad \lambda_1 < 0}\right. }
Its imaginary part is equal to
%of $Q(r)$ to vanish for all $r$
\eqn\Qim{{\rm{Im}} X(r)=\left({|M_{\pm}(r)|^{{2\over 3}}+G \over |M_{\pm}(r)|^{{1\over 3}}}\right)\times
\left\{\eqalign{&\sin{\left({2n\pi\over 3} \right)}\qquad\qquad\quad\,\, \lambda_1> 0\cr
&\sin{\left({\pi+2n\pi\over 3} \right)}\qquad\quad\,\,\lambda_1<0}\right.}
When we require the imaginary part to vanish for all $r$ we find that for positive values of $\lambda_1$ only
the solution with $n=0$ is meaningful. On the other hand, when $\lambda_1$ is negative the $n=1$ cubic root is
appropriate.

Let us now examine the position of the horizon, \ie, the roots of $X(r)=0$.
It is convenient to express $X(r)$ in the form
\eqn\Keq{X(r)={K(r)+K(r)^2-G\over K(r)}}
with $K(r)$ the real, same sign, cubic root of $M_{\pm}(r)$. From eq. \Keq\ we see that the solutions of $X(r)=0$ correspond to those
of a simple quadratic equation for $K(r)$ in terms of $G$. The discriminant $D=1+4G$ is always positive when $\lambda_2>\lambda_1^2$
so there are two solutions for $K(r)$
%\foot{In practice there is only one solution since $K(r)$ is always positive for $\lambda_2>\lambda_1^2$ and
%one of the two solutions is negative.} for $K(r)$. Namely,
\eqn\Kpm{K_{\pm}=-{1\over 2}\pm {1\over 2} \sqrt{1+4 G}}
Notice that $K_{+}$ is positive while $K_{-}$ is negative. Since $K(r)$ is either positive
or negative (depending on the sign of $\lambda_1$) for any $r$ when $\lambda_2>\lambda_1^2$,
only one of the solutions in \Kpm\ makes sense in each case.
Substituting either expression for $K_{\pm}$ in terms of $J(r)$ and $G$ into \Kpm\ and solving
for $J(r)$ yields
\eqn\Jrnot{J=-{1\over 2}\left(1+3G\right)}
Using then the definition of $J(r)$ from eq. \JGsol\ one finds that the position
of the horizon is $r=r_{+}$.

Next, we determine under which conditions $f(r)$ is positive for $r\geq r_{+}$.
%Recall that $K(r)$ is always positive since $G>0$.
It is easy to see that $X(r)$ is positive for $K(r)>K(r_{+})$ and $\lambda_1\geq 0$ while it
is negative for negative $\lambda_1$ and $K(r)<K(r_{+})$. To determine the sign of $X(r)$ outside
the horizon it is necessary to understand the monotonicity properties of $K(r)$.
Since
\eqn\monotonicityK{{\p K\over\p r}=\left\{\eqalign{ &{1\over 3} K {1\over \sqrt{J(r)^2+G^3}} {\p J\over\p r}\quad \lambda_1\geq 0\cr
                                               -&{1\over 3} K {1\over \sqrt{J(r)^2+G^3}} {\p J\over\p r} \quad \lambda_1<0}\right. \qquad\&\qquad
                  {\p J\over\p r}={3\over 2}{\lambda_2^2\over\lambda_1^3}{r_{+}\over r^2}}
$K(r)$ is a monotonically increasing function of $r$ when $\lambda_1>0$ and monotonically decreasing
when $\lambda_1<0$. It follows that the sign of $X(r)$ outside the horizon is the same as the sign
of $\lambda_1$ and given that $\lambda_2$ is positive, $f(r)$ is also positive.

In summary, a black hole solution with horizon located at $r=r_{+}$ exists for every
\eqn\ps{\lambda_2>\lambda_1^2\geq 0}

\item {(2)} Case II:$\quad G=0\Leftrightarrow \lambda_2=\lambda_1^2$

\noindent To examine this case it is easier to substitute $\lambda_2=\lambda_1^2$ directly in the equations
of motion. Taking the limit $\lambda_2\rightarrow \lambda_1^2$ in \bhsol\ yields the same result.
$X(r)$ is now given by
\eqn\Qrsp{X(r)=1+\left[2 {\cal{J}}(r)\right]^{{1\over 3}}\quad\&\quad {\cal{J}}(r)=-{1\over 2}+{3\over 2}\lambda_1\left(1-{r_{+}^6\over r^6}\right) }
%where ${\hat{J}}(r)$ denotes the limit of the function $J(r)$ defined in eq.\JGsol\ for $\lambda_2\rightarrow\lambda_1^2$.
It is easy to see that $X(r)$ is real as long as the real cubic root of ${\cal{J}}(r)$ is chosen.
Solving the equation $X(r)=0$ for ${\cal{J}}(r)$ determines the position of the horizon to be $r_{+}$ again.
Positivity of $f(r)$ outside the horizon is guaranteed since ${\cal{J}}(r)$ is a monotonically increasing (decreasing)
function of $r$ for $\lambda_1$ positive (negative).

Note however that ${\p X\over\p r}$ diverges at the point $r_0^6=\left({3\lambda_1\over 3\lambda_1-1}\right) r_{+}^6$ where
${\cal{J}}(r)$ vanishes. The same is true for the scalar curvature of the solution.
Nevertheless, if $r_0$ is negative or equivalently $0<\lambda<{1\over 3}$, the black hole solution remains valid.
This is also the case if the divergence occurs behind the horizon, in other words whenever $\lambda_1<0$.
On the other hand, for $\lambda_1>{1\over 3}$ and $\lambda_2=\lambda_1^2$ no consistent black hole solution exists.
When $\lambda_2=\lambda_1^2=0$ we recover the usual AdS Schwartzchild black hole of Einstein gravity.
At the point $\lambda_2=\lambda_1^2={1\over 9}$ the theory degenerates to pure Chern-Simons\foot{In fact the symmetries
of the theory are enhanced at the points $\lambda_2=-{2\over 9}+\lambda_1\pm {2\over 9}\left(1-3\lambda_1\right)^{{3\over 2}}$.
We thank Jose Edelstein for bringing this to our attention.}.

\item {(3)} Case III:$\quad G<0\Leftrightarrow \lambda_2<\lambda_1^2$

\noindent In this case  $\Delta(r)=J^2(r)+G^3$ is positive or negative depending on the value of the radial
coordinate $r$. To facilitate the analysis we introduce the coordinate $z=\left({r_{+}\over r}\right)^6$
and express $J(z)$ as
\eqn\Jz{J(z)=1-{3\over 2}{\lambda_2\over\lambda_1^2}+{3\over 2}{\lambda_2^2\over\lambda_1^3}\left(1-z\right)}
The horizon, if it exists, is now located at $z=1$. $J(z)$ is a monotonically
increasing function of $z$ when $\lambda_1$ is negative and decreasing when $\lambda_1$ is positive.
Notice that under this coordinate transformation eq. $\Delta(z)=0$ is a simple quadratic equation in $z$.
In particular, $\Delta(z)$ is positive for $J(z)>\sqrt{-G^3}$ and $J(z)<-\sqrt{-G^3}$ but negative otherwise.
%For simplicity we will analyze in detail only the case of positive $\lambda_1$, so $M(r)$ will be equal to $M_{+}(r)$
%in what follows.

Let us first consider what happens when $\Delta(z)$ is positive. Clearly, $M_{\pm}(z)$ is real and its cubic roots
can be expressed as $M_{\pm}(z)=|M_{\pm}(z)|^{{1\over 3}}e^{i {\chi_{\pm}+2n\pi\over 3}}$. One
can see that both $M_{\pm}$ are positive when $J(z)>\sqrt{-G^3}$ and negative when $J(z)<-\sqrt{-G^3}$.
We then set $\chi{\pm}=\pi$ for values of $z$ such that $J(z)<-\sqrt{-G^3}$ and $\chi_{\pm}=0$ for $J(z)>\sqrt{-G^3}$.
%for $M_{+}(z)$ when $J(z)>\sqrt{-G^3}$ while it is equal to $\pi$ when $J(z)<-\sqrt{-G^3}$. The same is true
%for $M_{-}(z)$, however for reasons that will be clear in the following we set $\chi=2\pi$
%It follows that
%\eqn\Qtwo{Q(z)=1+|M(z)|^{{1\over 3}} e^{i {\chi+2n\pi\over 3}}-{G\over |M(z)|^{{1\over 3}}}e^{-i {\chi+2n\pi\over 3}}}
%with the imaginary part equal to
%\eqn\Qimtwo{{\rm{Im}}Q(z)=\left({|M(z)|^{{2\over 3}}+G \over |M(z)|^{{1\over 3}}}\right)\sin{\left({\chi+2n\pi\over 3} \right)}}
Following the same reasoning as in case (I) we find that we must choose $n=0$ for values of $z$ such
that $J(z)>\sqrt{-G^3}$ and $n=1$ for those satisfying $J(z)<-\sqrt{-G^3}$.

On the other hand, when $z$ is such that $-\sqrt{-G^3}<J(z)<\sqrt{-G^3}$ the function $M_{\pm}(z)$ is complex and
equal\foot{Without loss of generality we choose here a positive imaginary part.} to $M_{\pm}(z)=J(z)\pm i \sqrt{-J(z)^2-G^3}$.
It can also be written as $M_{\pm}(z)=|M|e^{\pm i\phi(z)}$ with $\phi\in [0,\pi]$ and $|M|=\sqrt{-G^3}$.
Then $X(z)$ is
\eqn\Qthree{X(z)=1+|M|^{{1\over 3}} e^{i {\pm\phi+2n\pi\over 3}}-{G\over |M|^{{1\over 3}}}e^{-i {\pm\phi+2n\pi\over 3}}}
and Im$X(z)$ vanishes identically for any $n$. As a result, all three solutions are real in this region and
\eqn\Qreal{X(z)=1+2 \sqrt{-G}\cos{\left({\pm\phi+2n\pi\over 3}\right)} }

Let us summarize. Denote by $(z_{+},z_{-})$ the roots of $\Delta(z)=0$
\eqn\zab{z_{+}={2\lambda_1^3-3 \lambda_1\lambda_2+3\lambda_2^2+2 \left(\lambda_1^2-\lambda_2\right)^{{3\over 2}}\over 3\lambda_2^2}\quad\&\quad
z_{-}={2\lambda_1^3-3 \lambda_1\lambda_2+3\lambda_2^2-2 \left(\lambda_1^2-\lambda_2\right)^{{3\over 2}}\lambda_1^3\over 3\lambda_2^2}}
It is easy to see that $z_{+}>z_{-}$ for all $\lambda_2<\lambda_1^2$.
Depending on whether $\lambda_1$ is positive or negative $J(z_{+})$ is equal to the negative or positive value of $\pm\sqrt{-G^3}$.
%Note further that $z_1>z_2$ for all $\lambda_2\leq \lambda_1^2$
%\foot{To see this one needs
%the explicit expressions for $z_{1,2}$ in terms of $\lambda_1,\lambda_2$; \ie,
%\eqn\zab{z_1={2\lambda_1^3-3 \lambda_1\lambda_2+3\lambda_2^2+2 \left(\lambda_1^2-\lambda_2\right)^{{3\over 2}}\over 3\lambda_2^2}\quad\&\quad
%z_2={2\lambda_1^3-3 \lambda_1\lambda_2+3\lambda_2^2-2 \left(\lambda_1^2-\lambda_2\right)^{{3\over 2}}\lambda_1^3\over 3\lambda_2^2}}
%Combining this with $\lambda_2\leq \lambda_1^2$ for real values of the Lovelock parameters leads to $z_1\leq z_2$.}.
Bearing in mind that the monotonicity properties of $J(z)$ also depend on the sign of $\lambda_1$ we deduce that
\eqn\Qallp{X(z)=\left\{\eqalign{&1-|M_{+}(z)|^{{1\over 3}}+{G\over |M_{+}(z)|^{{1\over 3}}} \qquad\qquad z>z_{+} \cr
                 &1+2 \sqrt{-G}\cos{\left({\phi+2n\pi\over 3}\right)} \qquad z_{-}<z<z_{+} \cr
                 &1+|M_{+}(z)|^{{1\over 3}}-{G\over |M_{+}(z)|^{{1\over 3}}} \qquad\qquad z<z_{-} }\right.}
for positive $\lambda_1$ and
\eqn\Qalln{X(z)=\left\{\eqalign{&1-|M_{-}(z)|^{{1\over 3}}+{G\over |M_{-}(z)|^{{1\over 3}}} \qquad\qquad z<z_{-} \cr
                 &1+2 \sqrt{-G}\cos{\left({-\phi+2n\pi\over 3}\right)} \qquad   z_{-}<z<z_{+} \cr
                 &1+|M_{-}(z)|^{{1\over 3}}-{G\over |M_{-}(z)|^{{1\over 3}}} \qquad\qquad z>z_{+} }\right.}
for $\lambda_1$ negative.
Notice that it is impossible to construct a black hole solution for $\lambda_1^2>\lambda_2$ which is continuous both
at $z=z_{+}$ and at $z=z_{-}$. Consider for instance the case of positive $\lambda_1$. From the analysis of the imaginary
part of $X(z)$ we see that $X^{+}(z_{+})=X^{-}(z_{+})$ requires $n=1$ whereas $X^{+}(z_{-})=X^{-}(z_{-})$ requires $n=0$.
The opposite is true when $\lambda_1$ is negative\foot{Note that the roots $z_{\pm}$
are exactly equal to $P(x_{\pm})$, \ie, the values of $P(x)$ defined in section 2 at the extrema.}.

There are a number of ways to circumvent this problem and built consistent black hole solutions.
The obvious one is to constrain the parameter space of $\lambda_1,\lambda_2$ so that one of the branches in \Qallp\
and \Qalln\ is behind the horizon. In other words, we must require that at least one of $z_{\pm}$ is greater than unity.
We will now examine this case in detail.

Suppose first that $z_{+}>1$ but $0<z_{-}<1$. These two inequalities can be simultaneously solved in the
region $\lambda_1^2>\lambda_2$ for either positive or negative $\lambda_1$. In the former case, \ie, of positive
$\lambda_1$, $X(z)$ is given by
\eqn\Qpone{X(z)=\left\{\eqalign{&1+2 \sqrt{-G}\cos{\left({\phi\over 3}\right)} \qquad\qquad\quad z_{-}<z<z_{+} \cr
                                &1+|M_{+}(z)|^{{1\over 3}}-{G\over |M_{+}(z)|^{{1\over 3}}} \qquad\qquad z<z_{-} }\right.}
Note that we have chosen $n=0$ for the two branches to smoothly connect outside the horizon. This solution becomes singular
at $z=z_{+}$. For $z=1<z_{+}$ to define an event horizon, it should be a solution of the eq. $X(z)=0$ with $X(z)$ given by
the first branch of \Qpone. However, for $\phi\in [0,\pi]$ the first branch is strictly positive for any value of $z$. As a result
there is no consistent black hole solution in this region of parameter space.

Things are different when $\lambda_1$ is negative. Then $X(z)$ can be written as
\eqn\Qnone{X(z)=\left\{\eqalign{&1-|M_{-}(z)|^{{1\over 3}}+{G\over |M_{-}(z)|^{{1\over 3}}} \qquad\qquad z<z_{-} \cr
                 &1+2 \sqrt{-G}\cos{\left({-\phi+2\pi\over 3}\right)} \qquad   z_{-}<z<z_{+} }\right.}
with a singularity at $z_{+}$ hidden behind an event horizon at $z=1$. To see that $z=1$ is indeed a solution of
the eq. $X(z)=0$ in the ``complex" branch of \Qnone, bring the eq. $X(z)=0$ in the form
\eqn\rnoteq{\phi+2\pi=3 \arccos{\left[-{1\over 2\sqrt{-G}}\right]}}
and take the cosine function on both sides using the identity $\cos{3 x}=4\cos^3{x}-3\cos{x}$.
This directly leads to eq. \Jrnot\ which implies $z=1$. Note however that the existence of a horizon
in this case requires $\lambda_2\leq {3\over 4}\lambda_1^2$. Otherwise, $\sqrt{-G}<{1\over 2}$ and
$X(z)$ in the ``complex" branch is guaranteed to be strictly positive.

To determine the sign of $X(z)$ outside the horizon it is convenient to consider the behavior of the
``real branch" (by continuity the same will be true in the ``complex" branch). Recall that it can be written
as in \Keq\ with $K(z)$ the real, same sign, cubic root of $M(z)$. From the previous analysis, we know that
$M(z)$ is always negative in this region implying the same for $K(z)$. Moreover, $K^2+K-G$ is always positive
since for $\lambda_2\leq {3\over 4}\lambda_1^2$ the discriminant $D=1+4 G$ is negative. It follows that
$X(z)$ outside the horizon is negative. Given that $\lambda_1$ is also negative, we deduce that $f(z)$
is positive for all $z<1$. In summary we find that a black hole solution with $X(z)$ given by \Qnone\ is
valid for $\lambda_1,\lambda_2$ which satisfy
\eqn\sola{\lambda_1<0 \quad\&\quad \lambda_2<\lambda_1-{2\over 9}-{2\over 9}(1-3\lambda_1)^{{3\over 2}} \quad
                   OR\quad \lambda_1-{2\over 9}+{2\over 9}(1-3\lambda_1)^{{3\over 2}} <\lambda_2<{3\over 4}{\lambda_1^2}   }

Another possible case is for both $z_{\pm}$ to be greater than one. This is actually possible only when
$\lambda_1$ is negative. Following a similar line of reasoning we can then show that a consistent black hole
solution with $X(z)$ equal to
\eqn\Qntwo{X(z)=1-|M_{-}(z)|^{{1\over 3}}+{G\over |M_{-}(z)|^{{1\over 3}}} \qquad z<z_{-} ,}
a singularity at $z=z_{+}$ and a horizon for $z=1$ exists as long as
\eqn\solb{\lambda_1<0 \qquad {3\over 4}\lambda_1^2<\lambda_2<\lambda_1^2}

%Notice however that it is impossible to satisfy the requirement of continuity of the solution both at $z=z_{+}$ and at $z=z_{-}$
%Clearly if \bhsol\ with $Q$ given by \Qallp\ and \Qalln\ is to represent a black hole solution, the function $Q(z)$ must be
%continuous and smooth at $z=z_{+}$ and $z=z_{-}$. Observe however that when $\lambda_1$ is positive, $Q^{+}(z_{+})=Q^{-}(z_{+})$
%requires $n=1$ whereas $Q^{+}(z_{-})=Q^{-}(z_{-})$ requires $n=0$. The opposite is true when $\lambda_1$ is negative.
%Is it then possible to have a black hole solution for $\lambda_2<\lambda_1^2$?

One might have thought that these possibilities exhaust the spectrum of black hole solutions for $\lambda_1^2>\lambda_2$.
This however is not true. Notice that the two distinct solutions of $\Delta(z)=0$ are not necessarily
positive. In fact, either or both $z_{\pm}$ can be negative. This implies that the three branches in
\Qallp ( \Qalln) may be reduced to two or one branch where the continuity\foot{Imposing $X^{+}(z_{+})=X^{-}(z_{+})$ is not sufficient
to ensure continuity of the solution. Nonetheless, it is not difficult to show that first and second derivatives
coincide as well.}
%-{2\lambda_2^2 r_{+}\over\lambda_1^3 G}\left({r_{+}\over r}\right)^5$.}
condition $X^{+}(z_{\pm})=X^{-}(z_{\pm})$ can be applied.

%The answer is in the affirmative.
%This implies
%\foot{Imposing $Q^{+}(z_1)=Q^{-}(z_1)$ is not sufficient
%to ensure continuity of the solution. Nonetheless, it is not difficult to show for example that
%$\left. {\p Q\over\p r}\right|_{r_1^{+}}=\left. {\p Q\over\p r}\right|_{r_1^{-}}=
%-{2\lambda_2^2 r_{+}\over\lambda_1^3 G}\left({r_{+}\over r}\right)^5$.}
%that unless the three branches in \Qallp ( \Qalln) are reduced to two (or one), the solution of eq. \eqmgnl\ for
%$\lambda_2<\lambda_1^2$ does not represent a black hole.
%Is it possible to reduce the number of branches?
%The answer is in the affirmative. Although eq. $\Delta(z)=0$ has two distinct solutions $z_{\pm}$ it is not necessarily
%true that both of them are positive. In fact, there exist regions of the parameter space $(\lambda_1,\lambda_2)$ for which
%either only one of the solutions is positive or both are negative. In the former case, the three branches in \Qallp\ (\Qalln ) are reduced
%to two while in the latter to a single one. Let us examine all the possibilities carefully.

Consider first the simpler case, where both roots $z_{\pm}$ are negative. Eqs. \Qallp\ and \Qalln\ are then reduced to
\eqn\casea{X(z)= \left\{\eqalign{&1+|M_{-}(z)|^{{1\over 3}}-{G\over |M_{-}(z)|^{{1\over 3}}}\qquad \lambda_1<0\cr
                         &1-|M_{+}(z)|^{{1\over 3}} +{G\over |M_{+}(z)|^{{1\over 3}}}\qquad \lambda_1> 0}\right\}=
                         {K(z)^2+K(z)-G\over K(z)} }
where $K(z)$ is defined as the real, same sign, cubic root of $M_{\pm}(z)$.
From the analysis of Case I, recall that a necessary condition for the existence of a horizon is $D=1+4 G\geq 0$.
Then $K(z)$ evaluated at the horizon is equal to $K(z=1)=-{1\over 2}\left(1-\sqrt{1+4 G}\right)$. Note however that
this is incompatible with the first branch of \casea\ since $K(z)$ is positive for all $z$ when $J(z)\geq \sqrt{G^3}$.
We must then reduce the parameter space region to ${3\over 4}\lambda_1^2\leq \lambda_2<\lambda_1^2$ with $\lambda_1\geq 0$
and only consider the second of branch of \casea.
%following the steps in Case I we find that
%$f(z)={r_{+}^2\over z^{1/3} L^2}{\lambda_1\over\lambda_2}Q(z)$ is always positive outside the horizon.
Now $K(z)$ is negative and monotonically increasing with $z$. It follows that $K(z)<K(z=1)$ when $z<1$
so that $K^2+K-G<0$ and therefore $X(z)$ carries the same sign with $\lambda_1$ outside the horizon.
This proves that $f(z)>0$ outside the horizon given that $\lambda_2>0$.

Combining the necessary conditions for a black hole solution to exist in this case,
\ie, $z_{\pm}<0$ and ${3\over 4}\lambda_1^2\leq \lambda_2<\lambda_1^2$ and $\lambda_1>0$,
we find that eq. \bhsol\ defines a black hole geometry (with the cubic roots taken to be the real ones) as long as
\eqn\conditionscasea{0<\lambda_1<{1\over 3} \quad\&\quad 0<-{2\over 9}+\lambda_1+{2\over 9}\left(1-3\lambda_1\right)^{{3\over 2}}<\lambda_2<\lambda_1^2}

Let us move on to the case where $z_{+}$ is positive. Recall that we must also specify whether $z_{+}$ is greater
or smaller than one, \ie, inside or outside the horizon. First let us consider the situation for $0<z_{+}\leq 1$ and $z_{-}<0$.
Simultaneously satisfying both inequalities requires $\lambda_1$ to be positive. Then \Qallp\ leads to
\eqn\Qalla{X(z)=\left\{\eqalign{ &1-|M_{+}(z)|^{{1\over 3}}+{G\over |M_{+}(z)|^{{1\over 3}}} \qquad\qquad z>z_{+}\cr
                 &1+2 \sqrt{-G}\cos{\left({\phi+2\pi\over 3}\right)} \qquad\qquad  z<z_{+} }\right.}
The horizon now lies in the ``real" branch and $f(z)$ can be shown to be positive in a similar manner as before.
Existence of a horizon requires ${3\over 4}\lambda_1^2\leq\lambda_2<\lambda_1^2$. When combined with
$0<z_{+}\leq 1$ and $z_{-}<0$ leads to
\eqn\ineqa{\eqalign{&0<\lambda_1<{8\over 27} \quad\&\quad
{3\over 4}\lambda_1^2\leq \lambda_2<-{2\over 9}+\lambda_1+{2\over 9}(1-3\lambda_1)^{{3\over 2}}  \cr
                     &{8\over 27}\leq\lambda_1<{1\over 3} \quad\&\quad
-{2\over 9}+\lambda_1-{2\over 9}(1-3\lambda_1)^{{3\over 2}} <\lambda_2<
-{2\over 9}+\lambda_1+{2\over 9}(1-3\lambda_1)^{{3\over 2}}   } }
Eq. \ineqa\ represents the necessary and sufficient conditions on the Lovelock parameters for a black hole
solution with $X(z)$ given by \Qalla\ to exist.

On the other hand when $z_{+}>1$ and $z_{-}$ is negative, $X(z)$ can be expressed by a single ``complex" branch
with singularity at $z=z_{+}$ hidden behind the horizon
\eqn\QA{X(r)=1+2 \sqrt{-G}\cos{\left({\phi+2n\pi\over 3}\right)} \qquad\quad z<z_{+},\,\lambda_1\neq 0,\,\phi\in[-\pi,\pi]  }
Existence of a horizon reduces the parameter space region further to $\lambda_1<{3\over 4}\lambda_1^2$. With arguments similar to
the ones previously used we can prove that $f(z)$ is positive outside the horizon. We therefore find again a
consistent black hole solution. The parameter space region satisfying the above necessary and sufficient conditions is
\eqn\ineqb{\eqalign{
&\lambda_1 < 0 \quad\&\quad -{2\over 9}+\lambda_1-{2\over 9}(1-3\lambda_1)^{{3\over 2}} < \lambda_2<0 \quad OR \quad
0<\lambda_2 < -{2\over 9}+\lambda_1+{2\over 9}(1-3\lambda_1)^{{3\over 2}}  \cr
&0<\lambda_1 <{1\over 4} \quad\&\quad -{2\over 9}+\lambda_1-{2\over 9}(1-3\lambda_1)^{{3\over 2}} \leq\lambda_2<0 \quad OR\quad
0<\lambda_2\leq {3\over 4}\lambda_1^2 \cr
                     &{1\over 4}<\lambda_1<{8\over 27} \quad\&\quad
-{2\over 9}+\lambda_1-{2\over 9}(1-3\lambda_1)^{{3\over 2}} <\lambda_2\leq {3\over 4}\lambda_1^2  } }
It is useful to note that the expressions in \QA\ are not valid when $\lambda_1=0$. In fact a black hole solution exists
only when $\lambda_2$ vanishes as well. Understanding the $\lambda_2\rightarrow 0$ limit is also subtle. Computing
$X(z)=1+2\sqrt{-G} \cos\left[{1\over 3} \arccos[{J(z)\over \sqrt{-G^3}}]+{2 n\pi\over 3}\right]$ for $\lambda_2=0$ we find that
a solution to $X(z)=0$ exists only for $n=1$. This is in agreement
with known results on Gauss-Bonnet gravity; the solution with $n=1$ is continuously connected to the Gauss-Bonnet black hole.
For $n=2$ we recover the asymptotically AdS Gauss-Bonnet solution with naked singularity.

%In this case, taking the limit $\lambda_2\rightarrow 0$
%we recover the known black hole solutions of Gauss-Bonnet gravity (Cai).

%We will separately consider the situation for positive and
%negative values of $\lambda_1$. In the former case, \Qallp\ leads to
%\eqn\Qalla{Q(z)=\left\{\eqalign{ &1-|M_{+}(z)|^{{1\over 3}}+{G\over |M_{+}(z)|^{{1\over 3}}} \qquad\qquad z>z_{+}\cr
%                 &1+2 \sqrt{-G}\cos{\left({\phi+2\pi\over 3}\right)} \qquad\qquad  z<z_{+} }\right.}
%The position of the horizon is again $z=1$. That $z=1$ is a solution of the eq. $Q(z)=0$ in the "real"
%branch, is shown before. That it is also true in the "complex" branch, is easy to see. Simply bring
%eq. $Q(z)=0$ in the form
%\eqn\rnoteq{\phi+2\pi=3 \arccos{\left[-{1\over 2\sqrt{-G}}\right]}}
%and take the cosine function on both sides using the identity $\cos{3 x}=4\cos^3{x}-3\cos{x}$.
%This directly leads to eq. \Jrnot\ which implies $z=1$. Note however that the existence of a horizon
%in this case requires $\lambda_2\leq {3\over 4}\lambda_1^2$. Otherwise, $\sqrt{-G}<{1\over 2}$ and
%$Q(z)$ in the "complex" branch is guaranteed to be strictly positive. This is to be contrasted
%with the "real" branch, where existence of a horizon requires ${3\over 4}\lambda_1^2\leq\lambda_2<\lambda_1^2$.

\newsec{Weyl anomaly and the correspondence between flux positivity and causality}

\noindent In this Section we compute the Weyl anomaly for the third order
Lovelock theory in $AdS_7$ and determine the values of $t_2$ and $t_4$ using
the results of \deBoerPN.
We then show that the positivity of energy flux condition precisely matches
the causality condition studied in Section 2.
To compute the anomaly, we generalize the analysis of \deBoerPN\
where the Weyl anomaly was computed for a CFT defined by
Gauss-Bonnet gravity in $AdS_7$, to the theory defined by \Lthree.
The description below will be minimal;
for details of this procedure the reader is encouraged to consult
\deBoerPN.
The starting point is the ansatz
\eqn\adsanz{  ds^2=L_{AdS}^2 \left({1\over 4\rho^2} d\rho^2+{g_{ij}\over\rho} dx^i dx^j \right) }
where
\eqn\gijexp{ g_{ij}=g_{ij}^{(0)}+\rho g_{ij}^{(1)}+\rho^2 g_{ij}^{(2)}+\OO(\rho^3)  }
is an expansion  in powers of the radial coordinate $\rho$.
One can now solve the equations of motions derived from \Lthree\ order by order in the $\rho$
expansion and  determine $g_{ij}^{(i)}, i=1,\ldots$ in terms
of $g_{ij}^{(0)}$.
The resulting expansion \gijexp\ is then substituted back into  $\sqrt{\det g}\, \LL$ and
the coefficient of $1/\rho$ term encodes the anomaly.

As in \deBoerPN\ we take the boundary metric to be of the form
\eqn\bdmetric{  g_{ij} dx^i dx^j = f(x^3,x^4) \left[ (dx^1)^2  +(dx^2)^2\right] +\sum_{i=3}^6 (dx^i)^2 }
and use Mathematica to solve the equations of motion order by order in $\rho$.
The leading non-trivial term in the equations of motion relates the value of the AdS radius with
the cosmological constant via eq. \lads.
The next to leading term in the equations of motion determines $g^{(1)}$,
which is the same as in the Einstein-Hilbert case \HS:
\eqn\gijone{    g^{(1)}_{ij}= -{1\over4}\left(R_{ij}-{1\over10} R g^{(0)}_{ij}\right)  }
The solution
for $ g^{(2)}_{ij}$ is more non-trivial and involves both $\lambda_1$ and $\lambda_2$.

Substituting \adsanz\ together with the solution \bdmetric\
into the action \Lthree\ and extracting the $1/\rho$ term in the integrand
we obtain an expression of the form
$\int d^6 x \sqrt{\det g^{(0)}} \AA_W$.
We then demand that the coefficient in front of every term  in the expression
\eqn\eqcoeff{ \AA_W-\sum_{i=1}^3 b_i I_i -\sum_{i=1}^7 c_i C_i =0 }
vanishes.
In eq. \eqcoeff\ the $I_i$ are the B-type anomaly terms composed out of the Weyl
tensor, and $C_i$ are the total derivative terms.
Both can be found in Appendix A of \BastianelliFTb.
This completely fixes $b_i$ and $c_i$.
Since we are interested in the values of $t_2$ and $t_4$,
it is sufficient to obtain the results for $b_2/b_1$ and $b_3/b_1$.
These expressions are still too big to be quoted here.
As in \deBoerPN\ we can use them, together with the free field theory results\foot{
It should be noted that the resulting expressions for $t_2$ and $t_4$ are of course valid in
the strongly interacting CFT defined by Lovelock gravity in $AdS_7$.
The use of the free theory parametrization for the $b_i$ and $\AA,\BB,\CC$ is a technical
trick which works because the relation is linear and there are three independent
types of free CFTs in six dimension which contain scalar, fermion and antisymmetric two-form fields.},
to determine the values of $t_2$ and $t_4$ which determine the angular distribution
of the energy flux
\eqn\eqenergy{\vev{\cal{E}}={\vev {\epsilon^{*}_{ik}T_{ik}{\cal{E}}(\hat{n})\epsilon_{lj}T_{lj} } \over
\vev {\epsilon^{*}_{ik} T_{ik}\epsilon_{lj}T_{lj}} }=
{q_0\over \Omega_4}\left[1+t_2\left({\epsilon^{*}_{il}\epsilon_{lj}n_i n_j\over \epsilon^{*}_{ij}\epsilon_{ij}}-{1\over 5}\right)+
t_4 \left( {|\epsilon_{ij}n_i n_j|^2\over \epsilon^{*}_{ij}\epsilon_{ij} }-{2 \over 35}\right) \right]}
where  $\Omega_4={8\pi^2\over 3}$.
More precisely, we use
\eqn\anomalyton{\eqalign{
     1&={-}\left({28\over3} n_s{+}{896\over3} n_f{+}{8008\over3}n_a\right) \cr
     {b_2\over b_1}&=\left({5\over3}n_s{-}32 n_f{-}{2378\over3} n_a\right) \cr
     {b_3\over b_1}&=\left(2 n_s{+}40 n_f{+}180 n_a\right)  \cr }
}
and eqs (4.9)--(4.11) in \deBoerPN\ to determine $t_2$ and $t_4$.
It turns out that upon the substitution of \lads, the resulting expressions are remarkably simple:
\eqn\ttotfour{  t_2=-10 {2 \lambda_1+3 a^2(a^2-1)\over \lambda_1+a^2(3 a^2-2)}\qquad\qquad t_4=0 }

In Section 2 we analyzed the propagation of a graviton with helicity 2
in the black hole background.
The absence of metastable states propagating with speed larger than the
speed of light gives rise to the constraint $C\leq 0$, or, using the explicit expression for
$C$ \Cexpd,
\eqn\cc{ {5\lambda_1+a^2 \left(9 a^2-8\right)\over \left[\lambda_1+a^2\left(3 a^2-2\right)\right]^2}\geq 0 }
We want to compare \cc\  with the condition  coming
from the positivity of the energy flux \eqenergy.
The corresponding
constraint is given by the first line in eq. (4.8) in \deBoerPN\
which can be written as
\eqn\posc{  5-t_2\geq 0   }
Here we used the fact that $t_4$ vanishes in the case of third order
Lovelock gravity.
%Notice that $t_4$ vanishes in Lovelock gravity as in Gauss-Bonnet.
%As we saw in \deBoerPN, positivity of the energy flux then requires $t_2$ to be bounded by $t_2\leq 5$.
%To compare the constraints on the Lovelock parameters from the positivity of
%energy flux with those coming from causality of the boundary theory, it is useful to
%define the function $R(\lambda_1,\lambda_2)=t_2-5$.
%must be negative for the dual CFT to be consistent.
Using \ttotfour\ we can write eq. \posc\   as
\eqn\posca{ {5\lambda_1+a^2\left(9 a^2-8\right)\over \lambda_1+a^2\left(3 a^2-2\right)}\geq0}
Note the remarkable similarity between eqs. \cc\ and \posca: the numerators in
both formulas are exactly the same.
To ensure a complete matching between the two results we only need
to show that the denominator in eq. \posca\ is always positive for the
black hole solutions.
Incidentally, there are $AdS$ solutions for which this denominator
can be negative.
This is because the space of all AdS solutions is parameterized by
all possible values of $\lambda_1$ and all possible positive values of $a^2$,
since the value of $\lambda_2$ can always be determined from \lads.

In Sections 2 and 3 we analyzed the conditions for black hole solutions to exist.
One can show that in the parameter region where the black holes
exist (see Fig. 1) the denominator in \posca\ is indeed always
positive.
In fact, it vanishes along the dashed red curve in Fig. 1, which
is the border of the parameter  space where black hole solutions
are allowed.
Another way to see this is to note that the denominator in \posca\
is proportional to (minus) the expression \den.
To verify this, one needs to substitute the expression for $\lambda_2$
\lads\ into \den.

\newsec{Discussion}

\noindent In this paper we investigated the relation between causality and positivity
of the energy flux for CFTs defined by Lovelock gravities in AdS
spacetimes.
For a generic Lovelock gravity in arbitrary spacetime dimensions
we formulate the condition of existence of asymptotically AdS
black holes.
We study the propagation of gravitons of helicity 2 in these backgrounds
and determine constraints on the Lovelock coefficients resulting
from causality of the boundary theory.
We then consider cubic Lovelock gravity in $AdS_7$ in more
detail, and write down the black hole solutions explicitly.
To compute the parameters $t_2$ and $t_4$ in terms of the Lovelock
coefficients $\lambda_1$ and $\lambda_2$ we perform a holographic computation
of the Weyl anomaly along the lines of \deBoerPN.
The coefficients of the B-terms in the anomaly determine the three point
functions of the stress energy tensor which are in turn related
to the values of $t_2$ and $t_4$ \deBoerPN.
We find exact matching between the causality constraint
and the corresponding energy flux positivity constraint:
for given values of  $\lambda_1$ and $\lambda_2$, as long as the
black hole solution exists, causality is preserved  whenever
the energy flux is positive.

We expect other helicities to match as well (see e.g . \refs{\BuchelMyers,\Hofman,\CamanhoVW,\BuchelSK}).
In fact, it would not be surprising if any Lovelock theory of
gravity in any dimensions will have the matching property.
Our results [see eqs. \Cexpd, \cc\ and \ttotfour, and subsequent discussion]
lead us to the following conjectured expression
\eqn\ttwoeq{t_2-(d-1)=-{d-1\over (d-2)(d-3)}{\sum_p p ((d-2)(d-3) + 2 d (p-1) ) \lambdah_p
\alpha^{p-1}\over \sum_p p\lambdah_p \alpha^{p-1}}   }
for a $p$-th order Lovelock theory in $d+1$ dimensions.
We also conjecture that $t_4=0$ in all of these theories.
Note that eq. \ttwoeq\ reproduces both the third order Lovelock
gravity result in seven dimensions \ttotfour\ and the Gauss-Bonnet
result in any dimensions (eq. (3.32) of \BuchelSK; see also eq. (6.35) of \CamanhoVW)
The causality condition \causalitycondition\ together with \den\
then matches precisely the energy flux positivity condition
for arbitrary $p$ and $d$.

It is natural to ask how far the correspondence goes, both
on the gravity side and on the field theory side.
Can it be that all higher derivative gravities share this
property?
At this stage we do not have a definite answer since
the black hole solutions are not known given that the equations
of motion are complicated.
However one should note that we only need an asymptotic behavior
of the black hole metric to probe causality violation.
It might be possible to obtain such an asymptotic behavior
in a generic higher derivative gravity.
It will be harder to establish the condition for the black hole
existence: as we have seen, this condition is crucial for the matching to
work.
Namely, the positivity of energy condition may be naively violated
precisely at the point where the black hole solution ceases to exist.
Understanding these issues is important if we are to understand
better the role higher derivative gravities play in AdS/CFT.

Another question is whether the correspondence between causality at
finite temperature and the positivity of energy flux can be
shown directly in the field theory.
The gravity calculation implies that it is the near boundary
behavior of the metric which is responsible for causality
violation in the boundary theory.
This seems to imply that by studying causality in the short-distance behavior of the
two-point function of the stress-energy tensor at finite temperature
we should be able to constrain the three-point functions.
It should be interesting to see whether this can be made precise.

One of the interesting results of this paper is the
existence of regions in the parameter space where AdS solutions
exist but black hole solutions do not.
This is unlike Einstein-Hilbert gravity
where existence of an AdS solution implies existence
of an asymptotically AdS black hole.
From the point of view of the AdS/CFT correspondence the Hawking-Page transition
between the black hole and thermal AdS space \refs{\WittenZW-\HawkingDH} can be mapped to
the Hagedorn transition in field theory.
Indeed, the exponential density of gauge-invariant states $\rho\sim \exp(\beta_H E)$
in large $N$ gauge theories implies that partition function diverges for temperatures $T>1/\beta_H$.
Of course  this divergence signifies
the transition between the phases with $\OO(N^0)$ and $\OO(N^2)$
degrees of freedom.
The black hole is the gravitational description of the latter, high
temperature phase.
Now the fact the the black holes do not exist for some values
of Lovelock parameters may imply that the gravity theory in this region
is in some sense dual to a field theory whose density of states
does not grow so fast with energy.
In fact, higher derivative corrections to the Einstein-Hilbert
action are in some cases associated with finite $N$ corrections \refs{\AharonyXZ-\FayyazuddinFB,\KatsMQ},
which are expected to smooth out the phase transition.

On the other hand, what happens in
Gauss-Bonnet gravity may very well be the case for any Lovelock theory.
That is, for those values of the Lovelock parameters for which black
holes do not exist, the vacuum AdS solutions may be unstable \BoulwareWK.
It will be interesting to investigate this direction further by studying
small fluctuations around the AdS vacuum.

We have obtained the three point functions of the stress
energy tensor in a six-dimensional CFT dual to the cubic Lovelock theory.
It is interesting that unlike flat space, the cubic Lovelock term contributes to
these three point functions.
This can be seen from \ttwoeq, which receives contributions from three and
two point functions.
The third order Lovelock coupling
$\lambdah_3$ enters both explicitly and through the AdS curvature scale $\alpha$.
Remarkably, the three point functions of the stress energy tensor
satisfy the same relation as the one imposed by supersymmetry,
since $t_4=0$.
Of course this was also the case in Gauss-Bonnet gravity \deBoerPN,
but the vanishing of $t_4$ in the $\OO(R^3)$ gravity theory
is more non-trivial.
Again, this poses a question of how special the Lovelock theories
are in the AdS/CFT correspondence.
Can it be that all these theories are holographically mapped
to CFTs which are in some way related to their supersymmetric
parents?
After all, all known theories  whose gravity sector is the
Eistein-Hilbert gravity in $AdS_5$ are related in some way
(e.g. through orbifolding or deformation) to $\NN=4$ super Yang-Mills.
All these theories indeed retain  $a=c$ as a feature.

Finally, let us discuss the implications of our results on the
viscosity/entropy bound.
It has been shown that inclusion of a Gauss-Bonnet term in
gravity violates \BriganteLMSYa\ the $\eta/s\ge1/4\pi$  bound \KSS.
Causality implies \BriganteLMSYa\ that there is still a lower bound
on $\eta/s$  whose value depends on
the dimensionality of the space; the minimal value $\eta/s\ge 1/4\pi \times 219/529$
happens for $d=8$ \deBoerPN.
A straightforward but lengthy computation reveals that the  $p$-th order Lovelock term  with $p>2$  does
not contribute to the viscosity/entropy ratio of the boundary theory. In other words, the value of $\eta/s$
of the dual theory is completely determined by the coefficient of the Gauss-Bonnet
term in the gravitational action.
After arriving at this result we noticed that it has already appeared in the
literature \refs{\Medved-\ShuAX}.
%Recall here that the black brane
%entropy is independent of all Lovelock parameters.
Hence, the viscosity
to entropy ratio computed in \BriganteLMSYa\ remains valid for arbitrary Lovelock
theories of gravity
\eqn\etaovers{{\eta\over s}={1\over 4\pi} \left(1-{2 d\over d-2}\lambda_1\right)}
The positivity of energy flux (or, equivalently, causality) constraints place
restrictions on the allowed values of $\lambda_1$.
In the six-dimensional CFT dual to cubic Lovelock theory the constraints are
\eqn\conthree{   5-t_2\ge0,\qquad 5+{3t_2\over2}\ge0,\qquad 5+3 t_2\ge0  }
One can determine the physical region of Lovelock parameters
by plotting the curves defined by \conthree\ and restricting
to the physical region in Fig. 1.
It is interesting that the curves defined by inequalities in \conthree\
all meet at the point $(\alpha=3,\lambda_1=1/3)$.
In the physical region (see Fig. 1) the maximal allowed value of $\lambda_1$
comes from the last inequality in \conthree.
It is defined by the condition that the function $\lambda_1(\alpha)$
defined by  $5+3 t_2=0$ has a maximum.
This happens at $\alpha=15/8$ giving rise to $\lambda_1=64/165$.
Clearly, the value of $\eta/s$ is negative for this value of $\lambda_1$.
It would be interesting to see if the theory develops some
pathology before getting to this point.

\noindent{\bf Note added:} After we submitted our paper to the Arxiv we learned
of the forthcoming paper by Xian O. Camanho and Jose D. Edelstein \CE\ which
partially overlaps with our results.

\bigskip
\bigskip
\bigskip
\noindent {\bf  Acknowledgements:}  We thank J. Edelstein, D. Kutasov, K. Papadodimas, L. Rastelli and  M. Rocek
for useful discussions. M. K. thanks the YITP at Stony Brook University for hospitality. This work
was partly supported by NWO Spinoza grant and the FOM program {\it String Theory and Quantum Gravity}.

\appendix{A}{The viscosity/entropy ratio of the finite temperature CFT dual to Lovelock gravity.}

\noindent Here we compute the viscosity to entropy ratio for any $p$-th order Lovelock theory of gravity
in $d+1$ dimensions.
Our results agree with those of \refs{\Medved-\ShuAX}. We will mainly follow the discussion in \BriganteLMSYa.
For this we introduce the definitions and notations below
\eqn\newvar{\eqalign{z=&{r\over r_{+}}\qquad \tom={\omega\over r_{+}}\qquad \tq={q\over r_{+}} \qquad \tf (z)={f(z)\over r_{+}^2}\qquad \tgamma={\gamma\over r_{+}^d} \cr
\tR&=\sum_p \lambdah_p \left({\tf\over z^2}\right)^p z^d \qquad \tQ=\sum_p p\lambdah_p \left({\tf\over z^2}\right)^{p-1} z^{d-2} }}
We can now express the equation of motion for $\varphi$ defined in \Fourier\ with respect to the new variable $z$ as
\eqn\eqomz{\left[\tT_2 \p_z^2\varphi(z)+(\p_z\tT_2)\p_z\varphi(z)+\tT_0\varphi(z)\right]=0}
where
\eqn\Tsz{\tT_2={d-3\over 2}z^2 \tf \left(\p_z\tQ\right) \qquad \tT_0={\tT_2\over a^2\tf^2}\tom^2-{1\over 2} \tf \left(\p_z^2 \tQ\right) \tq^2}
and $a^2={1\over\alpha}$ in accordance with the discussion below \Tidef.
Evaluating the action to quadratic order on the solution of the equation of motion (neglecting contact terms) yields
\eqn\actionwq{S=-{1\over 2}{a r_{+}^d\over l_p^3} {2\over (d-2)(d-3)}\int {d\omega dq\over (2\pi)^2} T_2(z)(\p_z\varphi)\varphi(z)|_{z\rightarrow\infty} }

To compute the shear viscosity of the dual theory we set $q=0$ and focus on the low frequency limit of \eqomz.
Let us first study the behavior of the solution close to the horizon $z=1$. Note that
\eqn\nearhor{\left.{T_0\over T_2}\right|_{q=0}={\tom^2\over a^2 \tf}\simeq {\tom^2\over d^2 a^2 (z-1)^2}+\cdots\qquad\qquad {\p_z T_2\over T_2}\simeq {1\over z-1}+\cdots }
which implies $\varphi\simeq (z-1)^{\pm {i\tom\over d a}}$ when $z\simeq 1$. Choosing the infalling boundary condition at the horizon
\refs{\SonSS\HerzogHS\Marolf-\SkenderisSVR} we express the solution of \eqomz\ in the small frequency limit as
\eqn\phiw{\varphi=J(k)\left({a^2\tf\over z^2}\right)^{-i{\tom\over d a}} \left(1-i{\tom\over d a} g(z)+\cdots\right)}
with $g(z)$ regular at $z=1$ and $J(k)$ the boundary source for the field $\varphi$.

It is easy to see that the large $z$ expansion of $g(z)$ is of the form
\eqn\glargez{g(z)={A\over z^d}+\cdots}
with $A$ a constant to be determined later.
This leads to the following near boundary behavior for $\varphi,\,\p_z\varphi$ and $\tT_2(z)$
\eqn\allb{\eqalign{\varphi(z) &\simeq  J(k)\left(1-i{\tom\over d a} {A+ a^2\tgamma\over z^d}\right) \cr
\p_z\varphi &\simeq J(-k){i\tom\over a}{A+a^2\tgamma\over z^{d+1}} \cr
\tT_2(z) &\simeq {(d-3)(d-2)\over 2}{z^{d+1}\over a^2\tgamma}  }}
Substituting \allb\ into \actionwq\ yields a boundary term (to leading order in $\tilde{\omega}$)
\eqn\visca{S=-{1\over 2}\int {d\omega dq\over (2\pi)^2} {a r_{+}^d\over l_p^3} J(-k){i\tom\over a} {A+a^2\tgamma\over a^2\tgamma}J(k)   }
The viscosity of the dual plasma is then equal to
\eqn\viscb{\eta=\lim_{\omega\rightarrow 0}{{\rm Im} G_{ret.}(\omega,q=0)\over \omega}={r_{+}^{d-1}\over l_p^3} {A+a^2\tgamma\over a^2\tgamma} }

The entropy density of black branes in Lovelock theories of gravity has been discussed in \refs{\Caia-\Caib} (see also \refs{\Oleaa\Oleab\Kofinasb-\Kofinasa} where the holographic renormalization procedure is employed).
Remarkably, it is independent of all Lovelock coefficients $\lambdah_p$ with $(p>1)$
\eqn\entropy{s=4\pi {r_{+}^{d-1}\over l_p^3}.}
The ratio of \viscb\ to \entropy\ then yields
\eqn\etatos{{\eta\over s}={1\over 4\pi}  \left({A\over a^2\tgamma}+1\right) }
The contribution from the $p$-th order Lovelock terms is hidden in the ratio ${A\over a^2\tilde{\gamma}}$. 
Note that the denominator is roughly the product of the black hole mass, $\tilde{\gamma}$, with the square of the
AdS radius factor, $a$. 
Our objective in the following will be to determine $A$.

%We are interested in studying the limit small $\omega$ and $q$. We set $q=0$ in the following.
%So to evaluate the solution to leading order in $\omega$ for $q=0$ we write (Myers)
%\eqn\gadef{\varphi(z)=\left({\tf a^2\over z^2}\right)^{ -{i \tom \over d a}} \left(1-{i\tom\over d a} g(z) +\cdots\right)}
%and solve the equation for $g_1(z)$.

Let us start by making the following coordinate transformation
\eqn\cotr{x={\tf\over z^2}\qquad P(x)\equiv\sum_p \lambdah_p x^p={1\over z^d}}
Note that the position of the horizon is now at $x=0$ and of the boundary at $x=\alpha$.
With these definitions we also have that
\eqn\conseq{\tQ={\p_x P\over P^{1-{2\over d}}} \qquad {\p z\over\p x}=-{1\over d}{\p_x P\over P^{1+{1\over d}}} }
We can therefore express eq. \eqomz\ in terms of $x,P(x),g(x)$ and expand to linear order in $\tom$. The result is
\eqn\eqmlineartom{H_2 g''(x)+H_1 g'(x)+H=0}
where
\eqn\Hdef{\eqalign{ H_2=& x (\p_x) P \left[ (2-d) (\p_x P)^2+  d P \p_x^2 P \right] \qquad
H=d(\p_x^2 P)\left[ (\p_x P)^2-2 P \p_x^2 P \right]+d\, P(x) \p_x P\p_x^3 P \cr
H_1&=-(d-2)(\p_x P)^3+d\, x (\p_x P)^2 \p_x^2 P -2 d\, x P (\p_x^2 P)^2+d\, P (\p_x P) \left(\p_x^2 P +x \p_x^3 P\right)}}
Interestingly this equation has a rather simple solution for $\p_x g$
\eqn\gp{\p_x g={c_1 (\p_x P)^2-d\, P(x) (\p_x^2 P)\over x \left[(2-d)(\p_x P)^2+d\, P(x) \p_x^2 P\right]}.}
Here $c_1$ is an integration constant which can be fixed by imposing regularity at the horizon.
Observe that in the vicinity of $x=0$ the solution behaves like
\eqn\ghor{\p_x g(x)\simeq {c_1 \lambdah_1^2-2 d \lambdah_0\lambdah_2\over (2-d)\lambdah_1^2+2 d\lambdah_0\lambdah_2}{1\over x}+\OO(x^0)}
Recall that $\lambdah_0=1,\,\lambdah_1=-1$ and $\lambdah_2=\lambda_1$ the coefficient of the Gauss--Bonnet term in the action.
Demanding that the solution $g(x)$ be regular at the horizon translates then to
\eqn\cone{c_1=2 d\lambda_1}

Finally we would like to determine the behavior of $g(x)$ in the vicinity of the boundary $x=\alpha$.
In particular, we wish to specify $A$ appearing in eqs \viscb\ and \etatos.
In other words, we are interested in the asymptotic behavior as a function of the original variable $z$.
It is thus convenient to replace $x$ with its near the boundary expression $x\simeq \alpha(1+{\tgamma\over  z^d})$
and expand \gp\ to leading order in $z$ as $z\rightarrow\infty$. From \gp\ we deduce that
$\p_x g\simeq c_2 + {c_3\over z^d}+\cdots$ with
\eqn\AB{\eqalign{c_2&={\sum_{n,m}^p n(m c_1-d(n-1))\lambdah_n\lambdah_m \alpha^{n+m-2}\over \sum_{n,m}^p n(m (2-d)+d(n-1))\lambdah_n\lambdah_m \alpha^{n+m-1} }={2 d\lambda_1\over (d-2)\alpha } \cr
c_3&=c_1 a^2\tgamma \left[{\sum_{n,m}^p n(m c_1-d(n-1))(n+m-2)\lambdah_n\lambdah_m \alpha^{n+m-2}\over \sum_{n,m}^p n(m c_1-d(n-1))\lambdah_n\lambdah_m \alpha^{n+m-2}}-\right.\cr
\qquad\qquad\qquad &\left.-{ \sum_{n,m}^p n(m (2-d)+d(n-1))(n+m-1)\lambdah_n\lambdah_m \alpha^{n+m-1}\over \sum_{n,m}^p n(m (2-d)+d(n-1))\lambdah_n\lambdah_m \alpha^{n+m-1}}\right]  }}
where in the last equality of the first line we used \cone.
Note however that
\eqn\gb{g(z)=\int dz (\p_x g)\left({\p x\over \p z}\right)=
\int dz \left(c_2+{c_3\over z^d}+\cdots\right)\left(-d {\tgamma\over z^{d+1}}+\cdots\right)
\simeq {2 d\lambda_1 a^2\tgamma \over (d-2) z^d}+\OO\left({1\over z^{2 d}}\right)}
so the specific value of $c_3$ is irrelevant for the leading order boundary asymptotics of $g(z)$.
We are now in position to identify $A$ as
\eqn\Adef{A={2 d\lambda_1 a^2\tgamma \over (d-2) }}
and substituting into \etatos\ arrive at
\eqn\etatosf{{\eta\over s}={1\over 4\pi}  \left(1-{2 d\lambda_1\over d-2}\right)}
As a result only the Gauss-Bonnet coefficient affects the viscosity to entropy ratio. Any dependence
on the black hole mass cancels out \refs{\Medved-\ShuAX}.
\footatend\vfill\supereject\immediate\closeout\rfile\writestoppt
\baselineskip=14pt\centerline{{\bf References}}\bigskip{\frenchspacing%
\parindent=20pt\escapechar=` \input refs.tmp\vfill\eject}\nonfrenchspacing
\end